\documentclass[fleqn,usenatbib]{mnras}
\usepackage[dvipsnames]{xcolor}
\usepackage{siunitx}
\usepackage{graphicx}	
\usepackage{amsmath}	
\usepackage{amssymb}
\usepackage{hyperref}
\usepackage{tabularx}
\usepackage{mathptmx}
\usepackage{txfonts}
\usepackage{academicons}
\definecolor{orcidlogocol}{HTML}{A6CE39}
\usepackage[T1]{fontenc}
\usepackage{longtable}
\usepackage{subcaption}
\usepackage[symbol]{footmisc}
\usepackage[justification=centering]{caption}
\usepackage{relsize}

\usepackage{lipsum}

\title[Dynamics of dense gas around W42-MME]
{Fragmentation and dynamics of dense gas structures in the proximity of massive young stellar object W42-MME}

\author[N.~K. Bhadari et al.]{
N.~K. Bhadari$^{1,2}$\thanks{Email: naval@prl.res.in, bhadrinaval@gmail.com},
L.~K. Dewangan$^{1}$,
L.~E. Pirogov$^{3}$,
A.~G. Pazukhin$^{3}$,
I.~I. Zinchenko$^{3}$,
\newauthor{A.~K. Maity$^{1,2}$}, and\
Saurabh Sharma$^{4}$
\\\\
$^{1}$Physical Research Laboratory, Navrangpura, Ahmedabad 380009, India.\\
$^{2}$Indian Institute of Technology Gandhinagar Palaj, Gandhinagar 382355, India.\\
$^{3}$Institute of Applied Physics of the Russian Academy of Sciences 46 Ul’yanov str., 603950 Nizhny Novgorod, Russia.\\
$^{4}$Aryabhatta Research Institute of Observational Sciences, Manora Peak, Nainital 263002, India.\\
}

\date{Accepted 2023 September 28. Received 2023 September 27; in original form 2023 May 10}
\pubyear{2021}

\begin{document}
\label{firstpage}
\pagerange{\pageref{firstpage}--\pageref{lastpage}}
\maketitle

\begin{abstract}
%

We present an analysis of the dense gas structures in the immediate surroundings of the massive young stellar object (MYSO) W42-MME, using the high-resolution (0$''$.31$\times$0$''$.25) ALMA dust continuum and molecular line data. We performed a dendrogram analysis of H$^{13}$CO$^{+}$ (4--3) line data to study multi-scale structures and their spatio-kinematic properties, and analyzed the fragmentation and dynamics of dense structures down to $\sim$2000 AU scale. Our results reveal 19 dense gas structures, out of which 12 are leaves and 7 are branches in dendrogram terminology. 
These structures exhibit transonic--supersonic gas motions (1$<\mathcal{M}<5$) with overvirial states ($\alpha_{\rm vir}\geq2$).
The non-thermal velocity dispersion--size relation ($\sigma_{\rm nt}-L$) of dendrogram structures shows a weak negative correlation, while the velocity dispersion across the sky ($\delta\mathit{V_{\rm lsr}}$) correlates positively with structure size ($L$).
Velocity structure function ($S_{2}(l)^{1/2}$) analysis of H$^{13}$CO$^{+}$ data reveals strong power-law dependencies with lag ($l$) up to a scale length of $\lesssim$ 6000 AU. 
The mass-size ($M-R$) relation of dendrogram structures shows a positive correlation with power-law index of 1.73$\pm$0.23, and the leaf L17 hosting W42-MME meets the mass-size conditions for massive star formation.
Blue asymmetry is observed in the H$^{12}$CO$^{+}$ (4--3) line profiles of most of the leaves, indicating infall.
Overall, our results observationally support the hierarchical and chaotic collapse scenario in the proximity of the MYSO W42-MME.


\end{abstract}
%

%
%
\section{Introduction} \label{sec:intro}

Massive stars ($M>$8 $M_{\odot}$) typically form in giant molecular clouds (GMCs), which are the largest known complexes of molecular gas and dust in our galaxy.
Unlike the low-mass stars, where clear evolutionary steps have been established through theory and observations \citep[e.g.,][]{Shu1987,McKee2007}, the evolution of young massive stars has no well-defined stages \citep{Motte2018}.
Recent theoretical and numerical simulations have shown that the mass of newly formed stars is determined not only by small-scale mass accretion but also by large-scale, filamentary mass inflow/accretion \citep[e.g.,][]{Padoan2020}, which agrees with recent multi-scale and multi-wavelength observations \citep[e.g.,][]{Yuan2017,Dewangan2022,Liu2022_atomsV,Li2022,Xu2023,Liu2023}.
The low efficiency in star formation \citep[around 1--2\% in a single free-fall time scale of a cloud; e.g.,][]{Myers1986,Krumholz2005,Vutisalchavakul2016,Kim2021} hints that the turbulence and magnetic fields slow down the gravitational collapse of molecular clouds \citep[e.g.,][]{Myers1988,Vazquez-Semadeni2000,Bergin2007,Hennebelle2019,Krumholz2019}.
Also, the profound work by \citet{Larson1981} and follow up studies \citep[e.g.,][and references therein]{Solomon1987,Goodman1998,Bolatto2008,Heyer2009,Cen2021,Zhou2022} suggest that turbulence dissipates into the smaller regions of clumps and cores which are more prone to gravitational collapse leading subsequent star formation.
Consequently, the non-thermal motions on the scale of dense star-forming cores ($\sim$0.1 pc) can be sonic \citep[e.g.,][]{Myers1988,Caselli2002}, which agrees with the observations of filamentary, low-mass star-forming clouds \citep[][]{Hacar2011,Hacar2017,Pineda2015}.
However, the nature of turbulence in the massive star-forming clumps/cores remains ambiguous, as previous studies have shown a range of supersonic \citep[$\mathcal{M}\geq$2; e.g.,][]{Carolan2009} to subsonic \citep[$\mathcal{M}\leq$1; e.g.,][]{Li2020} non-thermal velocity dispersion.

Recent observations hint that the massive star formation (MSF) involves simultaneous onset of multiple physical processes from the scales of molecular clouds (>10 pc) to cores \citep[$\leq$0.1 pc;][and references therein]{Fukui2021,Hacar2022arXiv220309562H,Bhadari2022}. 
This may be possible because of the hierarchical fragmentation process (i.e., fragmentaion of cloud to high density structures of filaments, clumps and cores) that is primarily driven by the combined effect of turbulence and gravity \citep[e.g.,][]{Goodman2009,Inoue2013,Padoan2020,Fukui2021a,Fukui2021,Federrath2021}.
The gas motions of molecular clouds are coherent across scales of >10 pc to $\sim$0.1 pc, and exhibit a universal velocity dispersion-size scaling relation observed in different environments \citep[e.g.,][]{Solomon1987,Zhou2022,Liu2022}. 
This indicates that the dense fragments at different spatial scales are kinematically connected \citep{Rosolowsky2006,Rosolowsky2008,Goodman2009}.
In context to recent studies, the velocity dispersion-size relation appears to follow the \citet{Larson1981} first relation of $\Delta V \propto R^{0.38}$ with power law index in the range of 0.3--1.2 for larger scales (100--0.1 pc) \citep[e.g.,][]{Fuller1992,Goodman1998,Heyer2009,Falgarone2009,Cen2021,Izquierdo2021}, where $\Delta V$ and $R$ are the velocity dispersion and radius/size of cloud. 
The observed relation resembles the energy cascade in turbulent systems, indicating that turbulence dissipates from larger spatial scales to smaller ones.
The other two Larson's relations imply the virial equilibrium with the constant surface densities over all the spatial scales.
However, at the smaller scales of clouds where the density becomes high and multiple centers of collapse form, the role of both gravity and turbulence become important and hence at these scales the Larson's relations are debatable \citep[i.e., < 0.1 pc;][]{Ballesteros-Paredes2007, Ballesteros-Paredes2011,Traficante2018,Traficante2018larson}.
As a result, the effect of gravity and turbulence during the early stages of star formation including the massive ones, is one of the current hot topics in star formation research \citep[e.g.,][and references therein]{Ballesteros-Paredes2018,Federrath2021}.

\begin{figure}
	\begin{subfigure}{1\linewidth}
	\includegraphics[width=\textwidth]{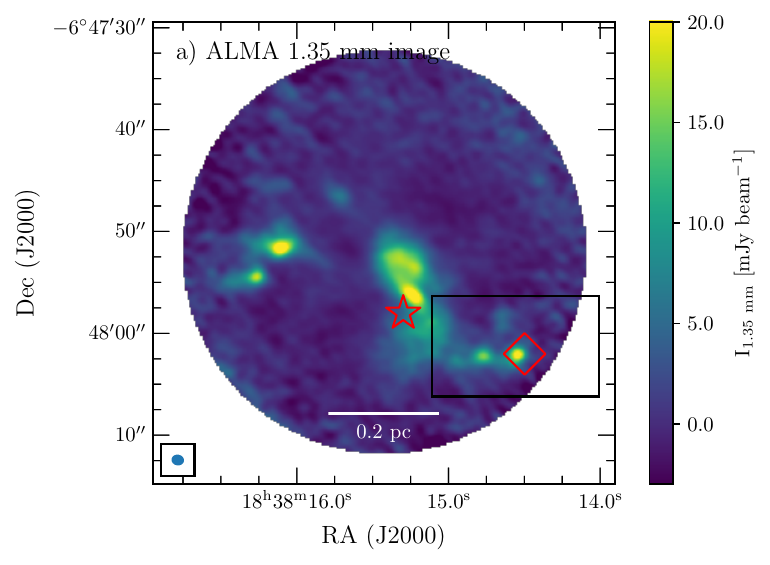}
	\end{subfigure}
\hfil
	\begin{subfigure}{1\linewidth}
	\includegraphics[width=\textwidth]{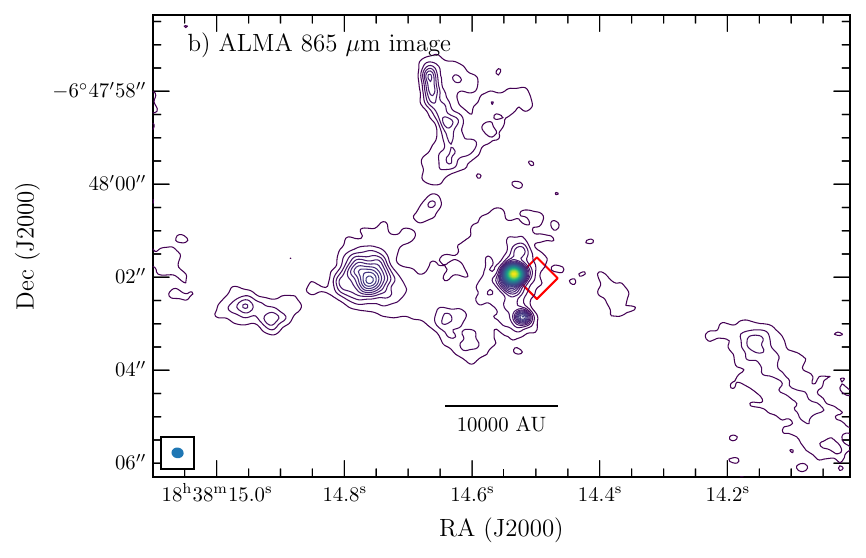}
	\end{subfigure}
\hfil
	\caption{a) ALMA 1.35 mm continuum emission map of an extended area around W42-MME. 
	b) ALMA 865 $\mu$m dust continuum contour map of an area marked by rectangular box in panel ``(a)''.
	The emission contour ranges from 0.4 to 43.55 mJy beam$^{-1}$ (i.e., maximum intensity of the image) in steps of 0.43 mJy beam$^{-1}$.
	The positions of the 6.7 GHz MME and ionizing O-type star are marked by diamond and star symbols, respectively. The scale bars are shown at $d=$ 3.8 kpc. The beam size of ALMA images are shown in bottom-left corner.}
	\label{fig:f1}
\end{figure}

The recent advancement of observational facilities such as the Atacama Large Millimeter/submillimeter Array (ALMA) has provided a great insights to the core scales in star-forming regions.
It extends our limit of studying the gas kinematics of dense structures form cloud to the core scales. 
This paper primarily makes use of the ALMA H$^{13}$CO$^{+}$ (4--3) data and presents a comprehensive study of the dense gas kinematics of a small area (0.304 $\times$ 0.181 pc$^{2}$) in W42 region \citep[distance $\sim$3.8 kpc; see][and references therein]{Dewangan2022}, known to host a bipolar H{\sc ii} region and Class~II 6.7 GHz methanol maser emission \citep[MME; see Figure~1 in][]{Dewangan2015a}. The H{\sc ii} region is powered by an O5-O6 star \citep[O5.5, hereafter;][]{Blum2000}, and the ionized and molecular gas have similar velocities of about 60 km s$^{-1}$, suggesting they are part of the same physical system \citep{Quireza2006,Anderson2009,Dewangan2015a}. The infra-red counterpart of 6.7 GHz MME in W42 (W42-MME, hereafter) is a rare young massive protostar (or massive young stellar object; MYSO hereafter), believed to be in an early evolutionary stage \citep{Dewangan2015b,Dewangan2022,Buizer2022}. It has a luminosity of approximately 4.5 $\times$ 10$^{4}$ $L_\odot$ and is located at the center of a parsec-scale bipolar outflow in H$_{2}$ image \citep[see][for more details]{Dewangan2015b}. 
Figure~\ref{fig:f1}a shows the ALMA 1.35 mm image of an extended area ($\sim$45$''$.4 $\times$ 45$''$.4) in the direction of W42-MME, which is further zoomed in to the 865 $\mu$m continuum image in Figure~\ref{fig:f1}b. The positions of W42-MME and the O5.5 star are also marked.
Recent high-resolution ($\sim$0$''$.3) molecular line study based on the ALMA data shows that the dusty envelope ($\sim$9000 AU) surrounding W42-MME hosts at least five continuum peaks \citep{Dewangan2022}. One of the major peaks associated with W42-MME shows the signature of bipolar outflow and a disk-like feature with velocity gradients. This study by \citet{Dewangan2022} suggests that MYSO W42-MME gains mass from its disk and the dusty envelope simultaneously.
In the larger physical extent of $\sim$3--5 pc, these authors found the presence of hub-filament system (HFS) which hosts our target region of study at its central hub-region \citep[$\sim$1--2 pc; see Figure~14 in][]{Dewangan2022}. 
It is to be noted that both the main sequence O5.5 star and the MYSO belong to the hub region. The radio continuum emission is also observed at 6 cm wavelength in the target area \citep[see Figure~1c in][]{Dewangan2022}. 
Targeting such region for high-resolution multi-wavelength and multi-scale study provides us a comprehensive understanding of different factors involved in MSF from the molecular cloud ($>5-10$ pc) to core scale ($<0.1$ pc).

In this paper we primarily utilized the high-resolution ALMA H$^{13}$CO$^{+}$ data to study the physical and kinematic properties of immediate surroundings of MYSO W42-MME, including the dusty envelope.
The paper is structured as follows: Section~\ref{sec:intro} serves as an introduction to the study, Section~\ref{data:sec} provides the description of data used in this study, and Section~\ref{results} outlines the results obtained from dendrogram analysis. 
Finally we discuss the consequences of the derived results in Section~\ref{sec:dis} and summarize the outcomes in Section~\ref{sec:conc}.

\section{Alma Data Sets}
\label{data:sec}

This paper utilizes H$^{12}$CO$^{+}$(4--3) ($\nu\sim$356.734 GHz), H$^{13}$CO$^{+}$(4--3) ($\nu\sim$346.998 GHz), and CH$_{3}$CCH (21$_{K}$--20$_{K}$) ($\nu\sim$358.709--818 GHz) data obtained from the ALMA in Band 7, acquired under project ID: \#2018.1.01318.S (PI: Lokesh Kumar Dewangan). The observed data has a beam size of 0$''$.31$\times$ 0$''$.25. The details of data acquisition and reduction are given in \citet{Dewangan2022}. Additionally, we have utilized the ALMA continuum data at 865 $\mu$m ($\nu\sim$346.5 GHz) from \citet{Dewangan2022}, which has a similar resolution to that of the line data. Furthermore, the publicly available ALMA continuum map at 1.35 mm (resolution $\sim$1$''$.2 $\times$ 1$''$.1) is obtained from the ALMA science archive (project ID \#2019.1.00195.L; PI: Molinari, Sergio). It is used to display the large area around W42-MME in Figure~\ref{fig:f1}a. The high-resolution Very Large Telescope (VLT) NAOS-CONICA (NACO) NIR adaptive-optics image at $K_{s}$ band ($\lambda$=2.18 $\mu$m; resolution $\sim$0$''$.2) is also utilized from \citet[][see more details in their paper]{Dewangan2015b}.

We used the ALMA H$^{12}$CO$^{+}$/H$^{13}$CO$^{+}$ data to analyze the dense gas kinematics and the CH$_{3}$CCH (21$_{K}$--20$_{K}$) data to estimate the temperature of the dense gas. In addition, we employed dust continuum maps to determine the mass of dense gas.

\section{Results}
\label{results}

\subsection{Dendrogram analysis of ALMA H$^{13}$CO$^{+}$(4--3)}
\label{dendro:analysis}

Dendrogram analysis is a powerful tool for studying the physical properties of hierarchical structure of star-forming molecular clouds. The dendrogram methodology, as described by \citet{Rosolowsky2008}, involves creating a tree diagram that characterizes emission structures based on their three-dimensional intensity isocontours. This method allows us to extract the emission structures at different spatial scales in a position-position-velocity ({\it p-p-v}) space, which is critical for understanding the complex interplay between dense gas kinematics, temperature, and star formation \citep[e.g.,][]{Shetty2012}.
The dendrogram is a hierarchical tree structure composed of two components: branches and leaves. Branches can split into multiple sub-structures (i.e., new branches and leaves), while leaves have no sub-structures. In this context, leaves (i.e., potential star-forming cores) represent the small-scale, bright structures at the tips of the tree that do not break down into further substructures. Branches (i.e., clumps), on the other hand, represent the large-scale, fainter structures lower in the tree that do break down into substructures.

To perform the dendrogram analysis, we utilized the python-based {\it astrodendro}\footnote{https://dendrograms.readthedocs.io/en/stable/index.html} package. This tool offers a method for determining the dendrogram structures within astronomical data in either 2D position-position ({\it p-p}) maps or 3D {\it p-p-v} data cubes. The algorithm {\it astrodendro} works by dividing the {\it p-p-v} data cube into a set of regions based on the distribution of intensity. The regions are then grouped together based on the specified similarity criterion, such as a threshold emission. This grouping process is repeated recursively until all of the regions are grouped together into a single structure. The result of this process is a tree-like diagram, or dendrogram, which represents the hierarchical structure of the cloud.

\subsubsection{Structure Identification and H$^{13}$CO$^{+}$ (4--3) moment maps} 
\label{str:iden}

\begin{figure*}
\includegraphics[width=1\linewidth]{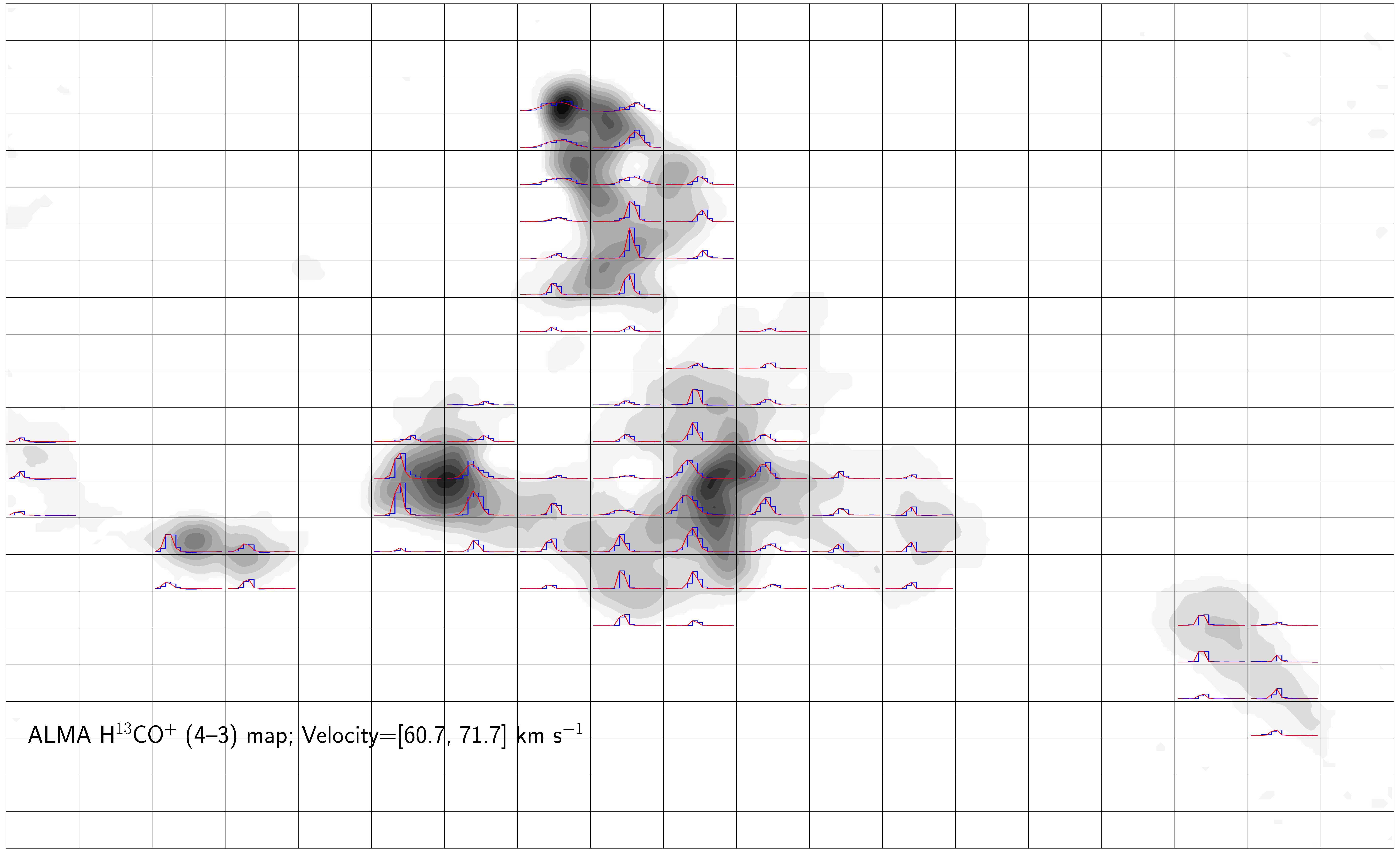}
\caption{ALMA H$^{13}$CO$^{+}$(4--3) moment-0 map overlaid with the averaged velocity profiles extracted along the regular grids of size $0''.864\times0''.432$. The extent of figure is identical to that of the Figure~\ref{fig:f1}b.
Profiles in red are the 1-D Gaussian fit to the observed spectra (in blue).
The y-scale ranges from $-$2.94 to 89.14 mJy beam$^{-1}$, while x-scale is $V_{LSR}$ from 60.73 to 71.70 km s$^{-1}$. 
} 
\label{fig1x}
\end{figure*}

\begin{figure}
\centering
\includegraphics[width=1\linewidth]{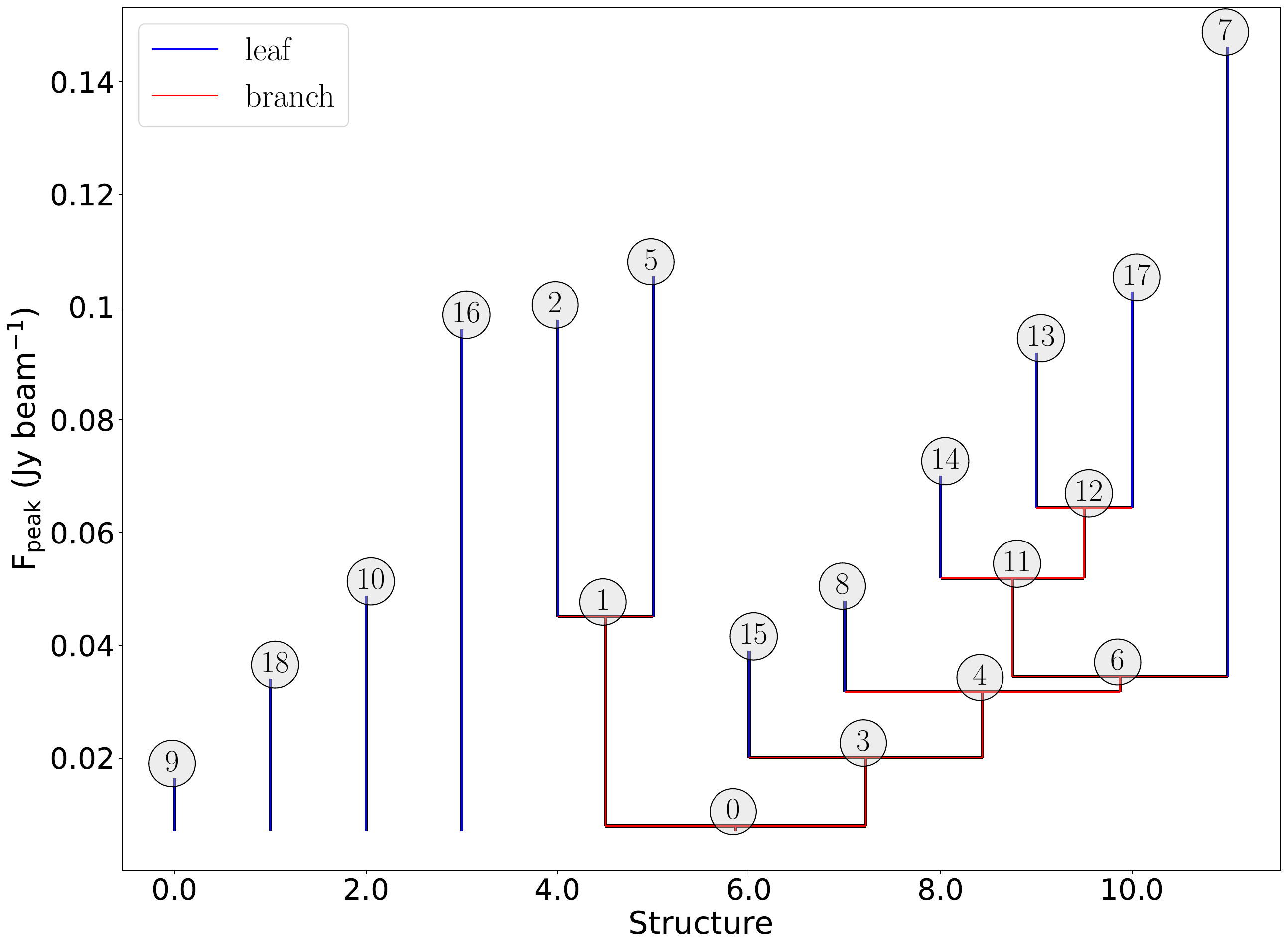}
\caption{Dendrogram tree of hierarchical structures identified from ALMA H$^{13}$CO$^{+}$(4--3) data. The leaves and branches are marked in the tree structure (see also  Figure~\ref{fig3x}).
} 
\label{fig2}
\end{figure}

We applied the dendrogram analysis to the ALMA H$^{13}$CO$^{+}$ (4--3) data (in {\it p-p-v} space) on 0.304 $\times$ 0.181 pc$^{2}$ area of our target region (see area footprint in Figure~\ref{fig:f1}b) in order to extract the hierarchical structure of the dense gas in the W42-MME region. 
The ALMA H$^{13}$CO$^{+}$ (4--3) line emission reveals mostly single peaked profiles toward the entire area of our target site, which is noticeable from Figure~\ref{fig1x} presenting the overlay of averaged H$^{13}$CO$^{+}$ spectra over the regular grids of $0''.864\times0''.432$ toward our target site. The background image is H$^{13}$CO$^{+}$ integrated intensity map (see following text in this Section).
The hierarchical tree diagram of dendrogram structures is extracted and is shown in Figure~\ref{fig2}. This exercise allowed us to identify and classify the different scales of the dense gas emission in the cloud, from the small-scale structures (i.e., $\lesssim0.01$ pc) to relatively larger structures($\geq0.05$ pc). Below are the steps we followed to identify the dendrogram structures.

To enhance the performance of the dendrogram algorithm, pixels with noise levels lower than three times the local noise level were masked out.
By analyzing the source/emission free individual velocity channels, we found that the local noise (or rms) varies from 1.6 to 3 mJy beam$^{-1}$ with mean value of $\sigma_{\rm rms}$=2.3 mJy beam$^{-1}$. 
The estimated $\sigma_{\rm rms}$ value is close to that of obtained by \citet{Dewangan2022} for a spectral resolution of 0.242 MHz.
We therefore masked out the noisy pixels which are below 3$\sigma_{\rm rms}$ (=7 mJy beam$^{-1}$) from the H$^{13}$CO$^{+}$ (4--3) data cube. 
This step is equivalent to set the ``min\_value'' parameter of {\it astrodendro} algorithm, which signifies the minimum emission in the Dendrogram tree structure. 
We further specified the values of other two crucial parameters used as inputs for the {\it astrodendro} algorithm. These are ``min\_delta'', and ``min\_npix''. 
The parameter ``min\_delta'', that indicates the minimum difference between two intensity peaks such that they can be considered as separate structures, was set to be 1$\sigma_{\rm rms}$ (=2.3 mJy beam$^{-1}$). 
The another parameter ``min\_npix'' indicates the minimum number of spatial-velocity pixels required for a leaf structure to be considered as independent entity. We have chosen this value such that a dendrogram structure contains at least two synthesized beam of ALMA H$^{13}$CO$^{+}$ (4--3) data. 
This is essential in order to measure the structure size and line width \citep{Rosolowsky2006}.
Thus, ``min\_npix'' was set to be 74 pixels (1 pixel= 0$''$.048).
The algorithm ultimately identified 19 structures including 7 branches and 12 leaves. 
Figure~\ref{fig2} displays the branches and leaves structures in the dendrogram tree.

\begin{figure}
\begin{subfigure}{1\linewidth}
\includegraphics[width=\textwidth]{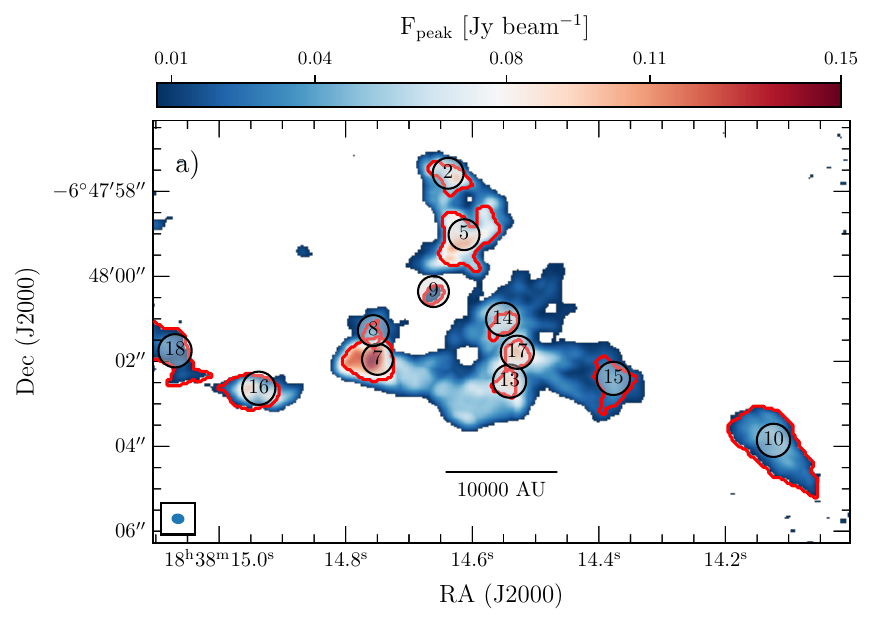}
\end{subfigure}

\begin{subfigure}{1\linewidth}
\includegraphics[width=\textwidth]{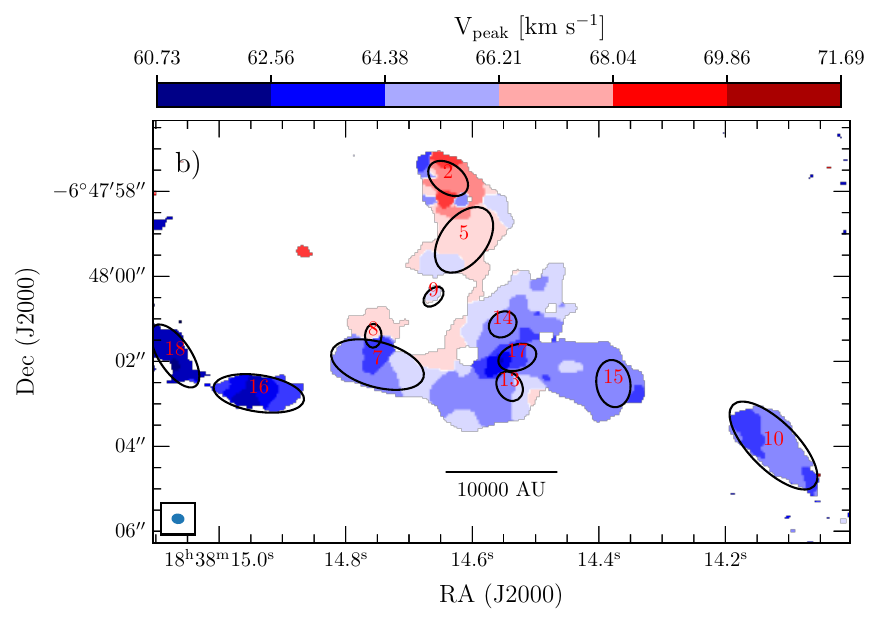}
\end{subfigure}

\begin{subfigure}{1\linewidth}
\includegraphics[width=\textwidth]{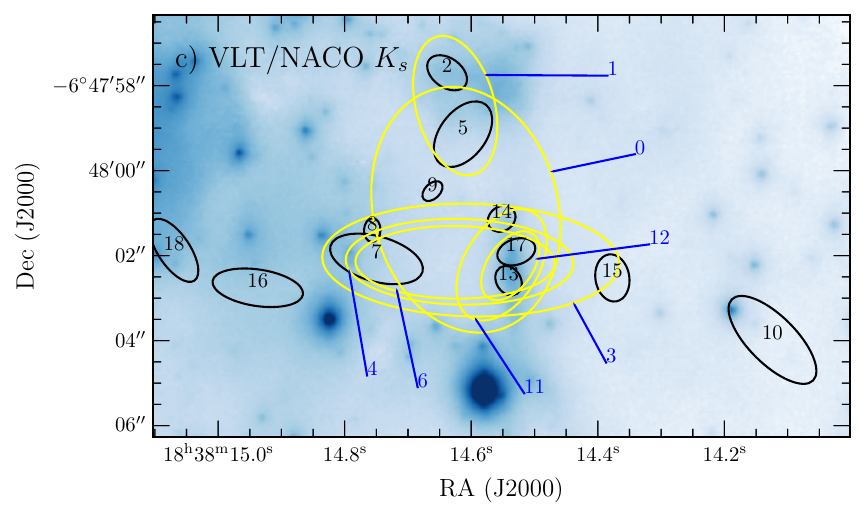}
\end{subfigure}

\caption{a) ALMA H$^{13}$CO$^{+}$(4--3) peak intensity map of region hosting W42-MME. 
The {\it astrodendro} identified leaf structures are overlaid and shown with contours and associated IDs, respectively.
(b) Peak velocity map derived using ALMA H$^{13}$CO$^{+}$(4--3) emission. 
c) VLT/NACO adaptive-optics $K_{s}$ band image.
In panels ``b'' and ``c'', the ellipses approximating the leaves (in black) and branches (in yellow) are shown. The structure ID for leaves and branches are displayed.
} 
\label{fig3x}
\end{figure}

We have superimposed the spatial distribution of the identified leaf structures on the peak intensity (F$_{\rm peak}$) map of the ALMA H$^{13}$CO$^{+}$ (4--3) data, as shown in Figure~\ref{fig3x}a. Each leaf structure is marked with its corresponding structure ID. The map provides a snapshot of the overall peak intensity distribution of the dense molecular gas. The dendrogram methodology can be understood by comparing the dendrogram tree structure (see Figure~\ref{fig2}) and the peak intensity map (see Figure~\ref{fig3x}a). 
Figure~\ref{fig3x}b presents the peak velocity (V$_{\rm peak}$) map of the ALMA H$^{13}$CO$^{+}$ (4--3) data which infers the gas velocity at peak intensity. A coherent velocity structure is noticeable along the $Dec\sim -06{\degr}47{''}02{'}$ at velocity of [62, 66] km s$^{-1}$. The northern structure, however, is seen at the redshifted velocities (see more discussion in Section~\ref{sec:dis}).
Figure~\ref{fig3x}b also shows the overlay of leaf structures on the V$_{\rm peak}$ map of H$^{13}$CO$^{+}$ emission, by their corrected sky-projected size. To infer the presence of point-like sources in our target area, we have displayed the VLT/NACO adaptive-optics $K_{s}$ band image of our target region in Figure~\ref{fig3x}c and overlaid the footprints of leaves and branches. 
As can be seen in Figures~\ref{fig3x}a--c (for comparison see Figure~\ref{fig2}), the central region is identified as dendrogram structure B0, which is separated from the other structures. This branch is further divided into 6 sub-branches and 8 leaves. Notably, leaf L17 belongs to the W42-MME, which has been previously described in \citep{Dewangan2022}.
Based on the VLT/NACO adaptive-optics $K_{s}$ band image, here we note that most of the point sources do not lie in the direction of identified structures. In other words, the turbulence in most of the dendrogram structures do not seem to be influenced by the presence of point-like sources.

\begin{figure}
\begin{subfigure}{1\linewidth}
\includegraphics[width=\textwidth]{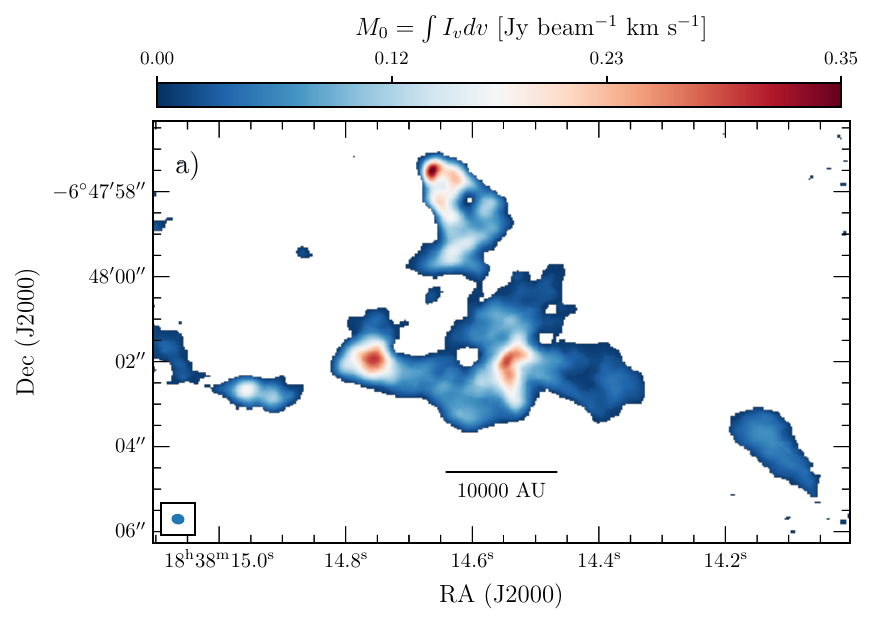}
\end{subfigure}

\begin{subfigure}{1\linewidth}
\includegraphics[width=\textwidth]{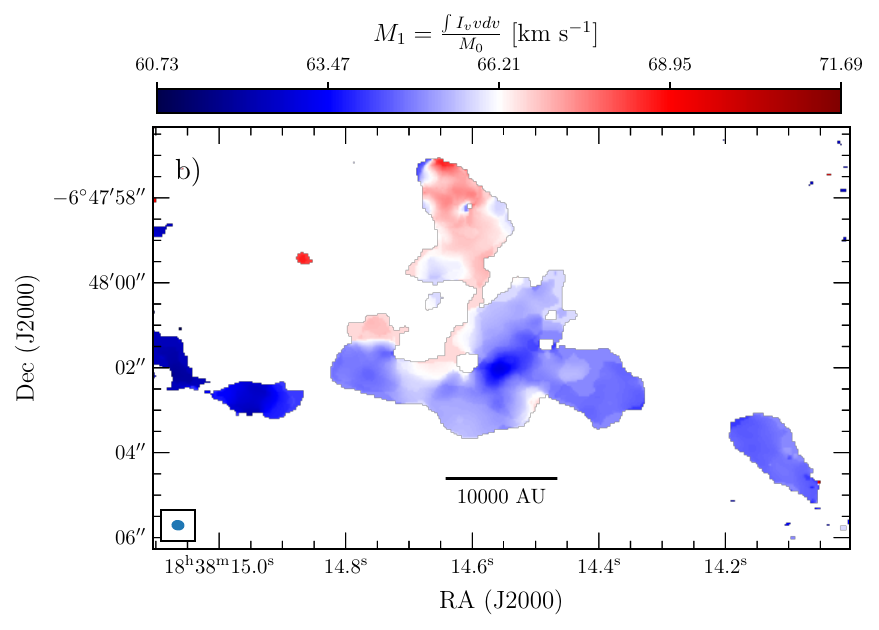}
\end{subfigure}

\begin{subfigure}{1\linewidth}
\includegraphics[width=\textwidth]{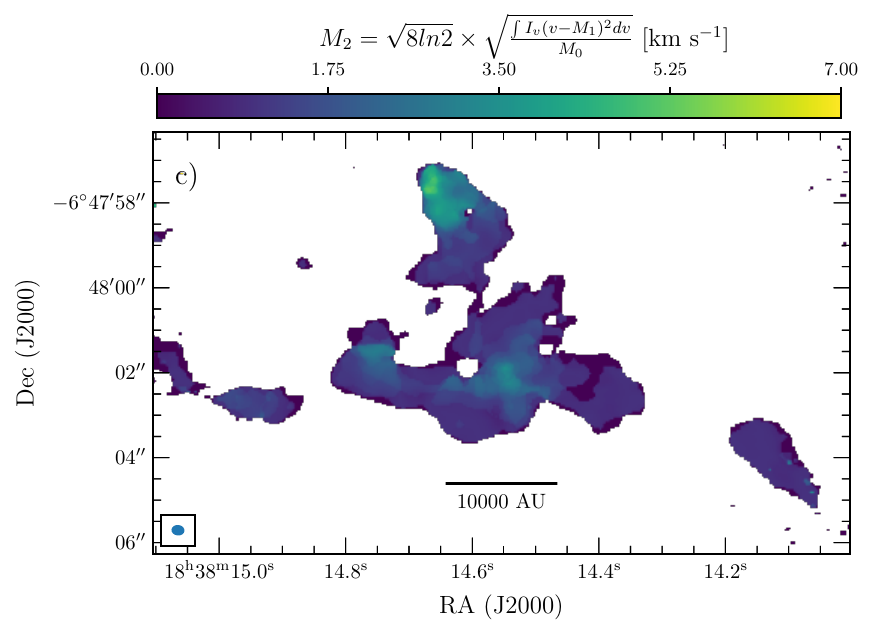}
\end{subfigure}

\caption{ALMA H$^{13}$CO$^{+}$ emission map of a) integrated intensity in the velocity range of [60.73, 71.70] km s$^{-1}$ (moment-0 map), b) intensity weighted velocity (moment-1 map) and b) intensity weighted FWHM velocity dispersion (moment-2 map) toward region hosting W42-MME.
} 
\label{fig3}
\end{figure}

We have also computed the moment statistics of ALMA H$^{13}$CO$^{+}$ (4--3) data. In millimeter--radio astronomy, moment maps provide a way to characterize the distribution of emission over a region of the sky. The first three moments i.e., moment-0; $M_0= \int I_{\varv} d \varv$, moment-1; $M_1=\frac{\int \varv I_{\varv} d \varv}{M_0}$, and moment-2; $M_2=\sqrt{8ln2}\times\sqrt{\frac{\int I_{\varv} (\varv - M_1)^2 d \varv}{M_0}}$, refer the integrated intensity over spectral/velocity axis, intensity weighted centroid velocity, and intensity weighted velocity dispersion (full width at half maximum (FWHM)), respectively.
Figure~\ref{fig3}a presents the $M_0$ map of ALMA H$^{13}$CO$^{+}$ emission in the direction of our target region, inferring the overall intensity/density distribution of dense gas. 
The velocity range of [60.73, 71.70] km s$^{-1}$ has been used for generating the $M_0$ map and identifying the dendrogram structures in {\it p-p-v} space. 
Figure~\ref{fig3}b shows the $M_1$ map providing the information of gas velocity distribution.
The $M_2$ map of extended area around W42-MME is shown in Figure~\ref{fig3}c, revealing the ALMA H$^{13}$CO$^{+}$ FWHM linewidth distribution.
The high value of velocity dispersion in the northern knot (leaf ID L2) is possible because of the presence of multiple velocity components (see blue spectra in Figure~\ref{fig1x}). However in the direction of W42-MME (leaf ID L17) and other leaves (ID L8 and L13), the high-dispersion value is observed regardless of the absence of multiple velocity components in H$^{13}$CO$^{+}$ spectra (see Figure~\ref{fig1x}).

\subsection{Physical parameters of dendrogram structures}
\label{str:phyparam}

The major and minor axis of dendrogram structures are related to {\it astrodendro} output parameters of ``major\_sigma" and ``minor\_sigma", which represent the 2nd spatial moments along longer and shorter direction, respectively. To obtain the physical extension of dendrogram structures (or major axis $\times$ minor axis), we need to multiply the outputs (i.e.,``major\_sigma" and ``minor\_sigma") by $\sqrt{\rm 8~ln2}$. We have also taken account of the filling factor ($f$) as discussed in \citet{Liu2022}. 
In other words, the {\it astrodendro} derived major and minor axis are less by a factor of $f=\sqrt{A_{\rm exact}/A_{\rm ellipse}}$, where $A_{\rm exact}$ and $A_{\rm ellipse}$ are the exact and calculated areas of Dendrogram structures in the sky plane \citep[see also][]{Rosolowsky2008, Liu2022}.

Table~\ref{tab:cores} lists the physical parameters of all identified structures, including structure ID, sky coordinates, integrated flux ($F^{\rm int}$), structure size (major axis($''$) $\times$ minor axis($''$)), structure length ($L$), position angle ($PA$; measured counter-clockwise from the RA axis), gas kinetic temperature ($T_{\rm kin}$), weighted mean velocity ($\langle V_{\rm lsr} \rangle$), velocity variation ($\delta V_{\rm lsr}$), observed weighted mean velocity dispersion ($\langle \sigma_{\mathrm{obs}} \rangle$), non-thermal velocity dispersion ($\langle \sigma_{\mathrm{nt}} \rangle$), three-dimensional (3D) Mach number ($\mathcal{M_{\rm 3D}}$), and the number of ALMA beams ($N$) that can fit within each structure. 
The estimation of $T_{\rm kin}$ is presented in Section~\ref{sec:temp}.
We have also computed the gas mass ($M_{\rm s}$), virial mass ($M_{\rm vir}$), virial parameter ($\alpha_{\rm vir}$), and gas density ($n$) of each dendrogram structure in Section~\ref{sec:mass}. 
\begin{table*}
\centering
\caption{Physical parameters of Dendrogram structures identified using ALMA H$^{13}$CO$^{+}$ data in W42 region.}
\label{tab:cores}
\resizebox{\textwidth}{!}{
\begin{tabular}{ccccccccccccccccccc}
\hline\hline
ID & $RA (J2000)$ & $Dec. (J2000)$ & F$_{\rm H^{13}CO^{+}}^{\rm int}$ & F$_{\rm 865~\mu m}^{\rm int}$ & ${\rm major \times minor}$ axis & $L$ & PA & $T_{\rm kin}$ & $\langle V_{\rm lsr}\rangle$ & $\delta V$ & $\langle\sigma\rangle$ & $\langle\sigma_{\rm nt}\rangle$ & $\mathcal{M_{\rm 3D}}$ & $M_{\rm s}$ & $M_{\rm vir}$ & $\alpha_{\rm vir}$ & $n$ & $N$ \\
  & hh:mm:sss  & dd:mm:sss & Jy & mJy &  ${\rm \arcsec  \times \arcsec}$  & ($\times 10^{3}$) AU & deg & K & km s$^{-1}$  &   km s$^{-1}$  &  km s$^{-1}$ & km s$^{-1}$ & & $M_{\odot}$ & $M_{\odot}$ &  & ($\times 10^{5}$) cm$^{-3}$ &  \\
\hline 

\multicolumn{11}{c}{\bf Leaves} \\

\hline
L2 & 18:38:14.64 & -6:47:57.66 & 0.76 & 10.72 & 1.05 $\times$ 0.7 & 3.24 & 144.67 & 64.28 $\pm$ 5.25 & 67.18 $\pm$ 0.01 & 0.37 $\pm$ 0.02 & 1.38 $\pm$ 0.02 & 1.37 $\pm$ 0.02 & 5.0 & 0.2 & 12.44 & 64.54 & 13.77 & 7.0\\
L5 & 18:38:14.61 & -6:47:59.14 & 1.5 & 22.18 & 1.77 $\times$ 1.1 & 5.27 & 52.26 & 61.84 $\pm$ 6.78 & 66.59 $\pm$ 0.02 & 0.3 $\pm$ 0.01 & 0.53 $\pm$ 0.02 & 0.52 $\pm$ 0.02 & 1.91 & 0.42 & 4.54 & 10.9 & 6.91 & 18.0\\
L7 & 18:38:14.75 & -6:48:02.05 & 2.61 & 76.41 & 2.27 $\times$ 1.07 & 5.91 & 164.47 & 45.02 $\pm$ 1.99 & 65.18 $\pm$ 0.02 & 0.45 $\pm$ 0.02 & 0.52 $\pm$ 0.02 & 0.51 $\pm$ 0.02 & 2.22 & 2.08 & 4.11 & 1.98 & 24.56 & 22.0\\
L8 & 18:38:14.76 & -6:48:01.37 & 0.08 & 4.46 & 0.57 $\times$ 0.4 & 1.79 & 87.28 & 64.02 $\pm$ 8.27 & 66.24 $\pm$ 0.04 & 0.65 $\pm$ 0.06 & 0.83 $\pm$ 0.05 & 0.82 $\pm$ 0.05 & 2.99 & 0.09 & 2.97 & 36.79 & 34.48 & 2.0\\
L9 & 18:38:14.66 & -6:48:00.47 & 0.03 & 1.78 & 0.57 $\times$ 0.36 & 1.72 & 223.34 & -- & 66.04 $\pm$ 0.13 & 0.23 $\pm$ 0.02 & 0.24 $\pm$ 0.1 & -- & -- & -- & -- & -- & -- & 2.0\\
L10 & 18:38:14.13 & -6:48:03.96 & 1.1 & 38.44 & 2.72 $\times$ 1.13 & 6.64 & 135.23 & 52.09 $\pm$ 11.81 & 64.62 $\pm$ 0.03 & 0.22 $\pm$ 0.01 & 0.34 $\pm$ 0.02 & 0.32 $\pm$ 0.03 & 1.28 & 0.88 & 3.05 & 3.47 & 7.31 & 28.0\\
L13 & 18:38:14.54 & -6:48:02.56 & 0.39 & 12.32 & 0.74 $\times$ 0.58 & 2.48 & 122.97 & 63.56 $\pm$ 2.79 & 64.96 $\pm$ 0.02 & 0.36 $\pm$ 0.03 & 0.85 $\pm$ 0.03 & 0.84 $\pm$ 0.03 & 3.08 & 0.23 & 4.46 & 19.88 & 35.64 & 4.0\\
L14 & 18:38:14.55 & -6:48:01.12 & 0.22 & 4.03 & 0.7 $\times$ 0.58 & 2.41 & 215.05 & 34.79 $\pm$ 8.92 & 65.16 $\pm$ 0.03 & 0.15 $\pm$ 0.01 & 0.49 $\pm$ 0.05 & 0.48 $\pm$ 0.05 & 2.39 & 0.15 & 1.69 & 11.22 & 26.3 & 4.0\\
L15 & 18:38:14.38 & -6:48:02.52 & 0.23 & 4.14 & 1.12 $\times$ 0.81 & 3.61 & 99.23 & 33.1 $\pm$ 2.84 & 64.67 $\pm$ 0.04 & 0.05 $\pm$ 0.01 & 0.4 $\pm$ 0.05 & 0.39 $\pm$ 0.05 & 1.96 & 0.17 & 1.79 & 10.88 & 8.5 & 8.0\\
L16 & 18:38:14.94 & -6:48:02.74 & 1.12 & 20.19 & 2.17 $\times$ 0.86 & 5.19 & 171.56 & -- & 63.37 $\pm$ 0.03 & 0.56 $\pm$ 0.02 & 0.44 $\pm$ 0.03 & -- & -- & -- & -- & -- & -- & 17.0\\
L17 & 18:38:14.53 & -6:48:01.87 & 0.65 & 67.05 & 0.93 $\times$ 0.61 & 2.85 & 198.2 & 69.71 $\pm$ 2.7 & 64.06 $\pm$ 0.01 & 0.41 $\pm$ 0.03 & 0.98 $\pm$ 0.02 & 0.97 $\pm$ 0.02 & 3.42 & 1.11 & 6.18 & 5.62 & 115.81 & 5.0\\
L18 & 18:38:15.07 & -6:48:01.84 & 0.24 & 5.19 & 1.73 $\times$ 0.76 & 4.34 & 124.16 & -- & 62.37 $\pm$ 0.04 & 0.34 $\pm$ 0.02 & 0.2 $\pm$ 0.03 & -- & -- & -- & -- & -- & -- & 12.0\\
\hline
\multicolumn{11}{c}{\bf Branches} \\ 
\hline
 B0 & 18:38:14.61 & -6:48:00.90 & 18.36 & 237.21 & 5.87 $\times$ 4.38 & 19.27 & 104.41 & 58.94 $\pm$ 3.87 & 65.8 $\pm$ 0.02 & 0.82 $\pm$ 0.01 & 0.43 $\pm$ 0.01 & 0.41 $\pm$ 0.01 & 1.56 & 4.71 & 13.75 & 2.92 & 1.6 & 234.0\\
B1 & 18:38:14.63 & -6:47:58.45 & 5.88 & 57.56 & 3.36 $\times$ 1.86 & 9.51 & 104.98 & 62.03 $\pm$ 3.27 & 66.66 $\pm$ 0.02 & 0.57 $\pm$ 0.01 & 0.66 $\pm$ 0.02 & 0.65 $\pm$ 0.02 & 2.41 & 1.08 & 10.53 & 9.78 & 3.05 & 57.0\\
B3 & 18:38:14.60 & -6:48:02.09 & 12.23 & 263.92 & 7.05 $\times$ 2.66 & 16.43 & 178.62 & 56.08 $\pm$ 2.04 & 65.33 $\pm$ 0.02 & 0.67 $\pm$ 0.01 & 0.43 $\pm$ 0.01 & 0.41 $\pm$ 0.01 & 1.58 & 5.55 & 9.49 & 1.71 & 3.04 & 170.0\\
B4 & 18:38:14.62 & -6:48:02.12 & 9.32 & 222.2 & 5.4 $\times$ 2.02 & 12.55 & 178.8 & 57.0 $\pm$ 2.44 & 65.44 $\pm$ 0.02 & 0.7 $\pm$ 0.01 & 0.52 $\pm$ 0.01 & 0.51 $\pm$ 0.02 & 1.96 & 4.59 & 9.11 & 1.99 & 5.64 & 99.0\\
B6 & 18:38:14.63 & -6:48:02.12 & 7.24 & 183.69 & 4.73 $\times$ 1.71 & 10.81 & 179.47 & 57.09 $\pm$ 3.71 & 65.48 $\pm$ 0.02 & 0.74 $\pm$ 0.01 & 0.56 $\pm$ 0.02 & 0.53 $\pm$ 0.02 & 2.05 & 3.79 & 8.26 & 2.19 & 7.27 & 74.0\\
B11 & 18:38:14.55 & -6:48:02.20 & 4.22 & 135.47 & 2.89 $\times$ 1.77 & 8.59 & 57.99 & 64.33 $\pm$ 3.52 & 64.92 $\pm$ 0.02 & 0.57 $\pm$ 0.01 & 0.62 $\pm$ 0.02 & 0.6 $\pm$ 0.02 & 2.19 & 2.44 & 8.98 & 3.69 & 9.34 & 46.0\\
B12 & 18:38:14.54 & -6:48:02.23 & 1.98 & 96.85 & 1.85 $\times$ 1.08 & 5.35 & 54.69 & 68.23 $\pm$ 2.62 & 64.53 $\pm$ 0.01 & 0.59 $\pm$ 0.02 & 0.81 $\pm$ 0.02 & 0.8 $\pm$ 0.02 & 2.81 & 1.63 & 8.18 & 5.03 & 25.86 & 18.0\\

\hline
\end{tabular}
}
\begin{flushleft}
\end{flushleft}
\end{table*}


We estimated the size of dendrogram structures by taking the geometric mean of corrected major and minor axes using the formula
\begin{equation}
\centering
 L (\rm AU)= D(\rm pc)~\sqrt{\rm major~axis('')\times minor~axis('')},
\end{equation}
where $D(=3.8 \times 10^{3})$ pc is the distance of W42. 
The $\langle V_{\rm lsr} \rangle$ represents the error-weighted mean velocity and is estimated using the $M_{1}$ map (see Figure~\ref{fig3}b) and the corresponding uncertainty map. Similarly, the weighted line-of-sight velocity dispersion/linewidth, $\langle \sigma_{\mathrm{obs}}\rangle$, is inferred from the $M_{2}$ map (see Figure~\ref{fig3}c) and the respective error map.
The uncertainties in the $M_{1}$ and $M_{2}$ maps are derived using Equations 5 and 6 from \citet[][not shown here]{Teague2019}. The final error in the mean of a parameter, denoted as $P$ (i.e., $V_{\rm lsr}$ or $\sigma_{\mathrm{obs}}$), is computed as :
%

\begin{equation}
\centering
  \Delta \langle P \rangle= \frac{1}{k}\sqrt{\sum_{i}^{k}\Delta p_{i}^{2}},
 \label{error}
\end{equation}
where $\Delta p_{i}$ and $k$ are the uncertainty in measurements of $P$ at $i^{th}$ pixel and the total number of pixels within the footprint of dendrogram structures, respectively.
The velocity variation, $\delta V_{\rm lsr}$, representing the plane-of-sky velocity dispersion, is estimated as the standard deviation of $V_{\rm lsr}$ within the structure. Its uncertainty is given by $\delta V_{\rm lsr}/\sqrt{2(k-1)}$. 
%


Additionally, we derived the non-thermal velocity dispersion as:
\begin{equation}
\centering
\langle \sigma_{\mathrm{nt}} \rangle = \sqrt{\langle \sigma^2_{\mathrm{obs}}\rangle - \sigma^2_{\mathrm{th}}},
\end{equation}
where $\sigma_{\mathrm{th}}$ is the thermal velocity dispersion for dendrogram structure, and is given by

\begin{equation}
\centering
\sigma_{\rm th} = \sqrt{\frac{k_{\rm B}~T_{\mathrm{kin},i}}{\mu m_{\mathrm H}}}.
\label{eq:thermal}
\end{equation}
In Equation~\ref{eq:thermal}, $\mu$=30 is the the molecular weight of the H$^{13}$CO molecule, $m_{\rm H}$ is the mass of atomic hydrogen (approximating to proton mass), $k_{\rm B}$ is the Boltzmann constant, and $T_{\mathrm{kin},i}$ is the gas kinetic temperature of the $i^{th}$ structure derived using CH$_{3}$CCH (21$_{K}$--20$_{K}$) emission (see Section~\ref{sec:temp}). The value of sound speed $c_{\rm s}$ in a medium is influenced by all the gas particles present in it. This can be determined using Equation~\ref{eq:thermal} with a mean molecular weight per free particle of $\mu$=2.37 \citep{Kauffmann2008}.
The 3D Mach number, $\mathcal{M_{\rm 3D}}$, is estimated by $\sqrt{3}\sigma_{\rm nt}/c_{\rm s}$.

\subsubsection{Dense gas temperature ($T_{\rm kin}$) estimation}
\label{sec:temp}
 
\begin{figure}
\begin{subfigure}{1\linewidth}
\includegraphics[width=\textwidth]{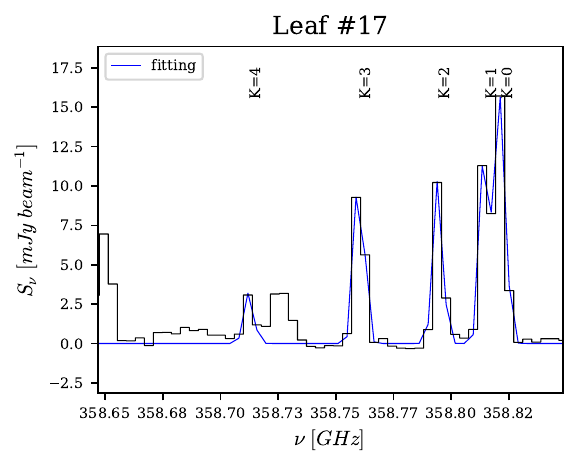}
\end{subfigure}

\begin{subfigure}{1\linewidth}
\includegraphics[width=\textwidth]{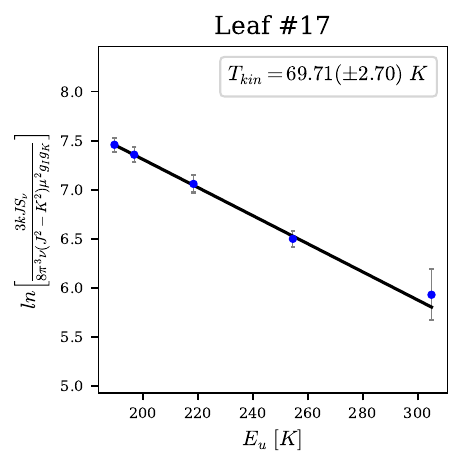}
\end{subfigure}

\caption{{\it Upper panel}: Averaged CH$_{3}$CCH (21$_{K}$--20$_{K}$) spectra (in black) and best fit model of CH$_{3}$CCH lines (in blue) for Dendrogram leaf L17.
{\it Lower panel}: CH$_{3}$CCH rotation diagrams of leaf L17 (see text for details). 
A best fit line to the 5 CH$_{3}$CCH transitions is shown and corresponding $T_{\rm kin}$ is labeled.
} 
\label{fig:rotdia}
\end{figure}

Previous studies show that the low excitation level (e.g., $J=$5--4) transitions from CH$_{3}$CCH molecule give a good estimate of the kinetic temperature at gas density of $\sim$10$^{3-4}$ cm$^{-3}$ \citep{Askne1984,Bergin1994} . However, the higher level transitions can trace gas density  $\ge$10$^{5-6}$ cm$^{-3}$ \citep[e.g.,][]{Aladro2011,Santos2022}, which can be used to estimate the gas temperature of dense environment as traced by  HCO$^{+}$ (4--3) transition with a critical density of \citep[i.e., $\sim$3.5$\times$10$^{6}$ to 2$\times$10$^{6}$ cm$^{-3}$ for the temperature range of 10--100 K, respectively;][]{Shirley2015}.
Therefore, we utilized the ALMA Band 7 CH$_{3}$CCH data to estimate the gas kinetic temperature ($T_{\rm kin}$) of dendrogram structures identified in H$^{13}$CO$^{+}$ (4--3) transition. The observed data contains five components of the CH$_{3}$CCH (21--20) K-ladder with K = 0-4, and we generated the corresponding averaged spectrum for each dendrogram leaf and branch structure. 
The rotational temperatures for the dendrogram leaf and branch structures are determined using these transitions. 
The spectrum and corresponding rotation diagrams of leaf L17 are presented in Figure~\ref{fig:rotdia}.
The diagram plots the values proportional to natural logarithm of the measured integrated line intensity, $S_{\nu}$, against the upper energy level $E_{u}$ of specified molecule.
The $y-$axis corresponds to a parameter defined by the quantum numbers $J$ and $K$, the transition frequency $\nu$, the dipole moment $\mu$, the degeneracy $g_{K}$ associated with the internal quantum number $K$, and the statistical weight $g_{I}$ associated with the nuclear spin.
The slope a of the fitted straight line, $y = ax+b$, is related to the kinetic temperature, $T_{\rm kin}= -\frac{1}{a}$, while the uncertainty in the estimation of $T_{\rm kin}$ is $\Delta T_{\rm kin}= \frac{\Delta a}{a^2}$. 

The temperature estimates for leaf structures ranged from 33 to 70 K, with the mean (median) values of 54.27 (61.84) K,  while for branches the range is [56.08, 68.23] K with mean (median) value of 60.53 (58.94) K.
It is important to note that in the case of leaves L9, L16, and L18, weak emission is observed in CH$_{3}$CCH (21--20) line, which made it difficult to define $T_{\rm kin}$ of these structures.

%


\subsubsection{Dense gas mass (M$_{\rm s}$) and virial parameter ($\alpha_{\rm vir}$) estimation}
\label{sec:mass}

The dust/gas mass can be estimated from the knowledge of the dust emission in the interstellar medium (ISM). 
With the assumption of spherical dust grains and the optically thin dust emission as a modified black body emission in the Rayleigh-Jeans limit, the gas mass can be estimated as:
\begin{equation}
 M_{\rm s}= \frac{F_\nu~D^2~R_t}{B_{(\nu,T_d)}~k_\nu},
\end{equation}
where $F_\nu$, $D$, $B_{(\nu,T_d)}$, $R_t$, and  $k_\nu$ are total integrated flux at observing frequency $\nu$, distance of object, Planck function at dust/gas temperature T$_d$, gas-to-dust mass ratio, and dust absorption coefficient, respectively.
%
%

By adding up the emission inside the boundary of each dendrogram structure, we calculated $F_\nu$ at $\nu=346.58$ GHz (or 865 $\mu$m) for each one.
We derived the $k_\nu$ by adopting a relation $k_\nu= 10\left(\frac{\nu}{1.2~({\rm THz})}\right)^{p}$, where $p=1.5$ is adopted for all the dense cores \citep[e.g.,][and references therein]{Bracco2017}. We therefore used $k^{865 \mu m}_{\rm 346.58~GHz}$= 1.55 cm$^{2}$ g$^{-1}$ for the mass estimation.
Using the derived gas temperatures in Section~\ref{sec:temp}, $D=3.8$ kpc, and $R_t=100$ \citep[][]{Weingartner2001}, we estimated the gas mass for each dendrogram structure, which is tabulated in Table~\ref{tab:cores}. 
The primary source of uncertainty in mass estimation is the dust emissivity, which is uncertain by a factor of a few.
We used the assumption of spherical symmetry for the dense cores that were identified and calculated their volume densities using the $M_{\rm s}$ and effective radius (i.e., $R_{\rm eff}=L/2$). The formula used for this purpose was $n_{\rm H_2} = \frac{M_{\rm s}}{(\frac{4}{3}\pi R_{\rm eff}^3) (2.8 ~m_{\rm H})}$. The volume densities for leaf (branch) structures varied from 3.49 (0.16)$\times 10^{6} {\rm cm^{-3}}$ to 12.75 (2.59)$\times 10^{6} {\rm cm^{-3}}$, with mean and medium values of 3.08 (0.8)$\times 10^{6} {\rm cm^{-3}}$ and 2.36 (0.56)$\times 10^{6} {\rm cm^{-3}}$.

We have also estimated the virial mass of identified structures using equation \citep[][]{Bertoldi1992,Singh_ayushi2021ApJ}
\begin{equation}
 M_{\rm vir}= \frac{5}{\beta_{1}\beta_{2}}\frac{\sigma_{\rm total}^{2}R_{\rm eff}}{G},
\end{equation}
where $\sigma_{\rm total}^{2}= \sqrt{\sigma_{\rm nt}^{2}+c^{2}_s}$ is the total velocity dispersion with the sound speed of $c_s$, and $G$ is the gravitational constant. The parameters, $\beta_{1}=\frac{(1-b/3)}{(1-2b/5)}$ belongs to a power-law density profile of $\rho\propto r^{-b}$ with $b=1.6$ \citep[e.g.,][]{Pirogov2009} and $\beta_{2}=\frac{sin^{-1}(e)}{e}$ is a geometry factor with eccentricity
$e=\sqrt{1-f^{2}_{\rm int}}$ of elliptical structures.
The intrinsic axis ratio ($f_{\rm int}$) is related to the observed axis ratio ($f_{\rm obs}$; see also Figure~\ref{fig4}) as \citep{Fall1983AJ,Li2013ApJ}
\begin{equation}
 f_{\rm int}= \frac{2}{\pi}~f_{\rm obs}~{\mathcal F_1(0.5, 0.5, - 0.5, 1.5, 1, 1-f^{2}_{\rm obs})},
\end{equation}
where $\mathcal F_1$ is the Appell hypergeometric function of the ﬁrst kind. The power-law density profile of $\rho\propto r^{-b}$ with $b=1.6$ is observed for the inner regions of the sample of 16 cores associated with MSF \citep[e.g.,][]{Pirogov2009}. For low-mass star-forming regions $b$ is usually higher.
The derived $\beta_{2}$ value of dendrogram structures varies from 1.1 to 1.4.

Based on the derived gas mass ($M_{\rm s}$) and virial mass ($M_{\rm vir}$) of structures, we estimated the virial parameter, $\alpha_{\rm vir}= \left(\frac{M_{\rm vir}}{M_{\rm s}}\right)$, that infers the dynamical state of gas structure. Table~\ref{tab:cores} lists the estimated mass and virial parameter of dendrogram structures.

\subsubsection{Spectral profiles of dendrogram leaves}

\begin{figure}
\includegraphics[width=1\linewidth]{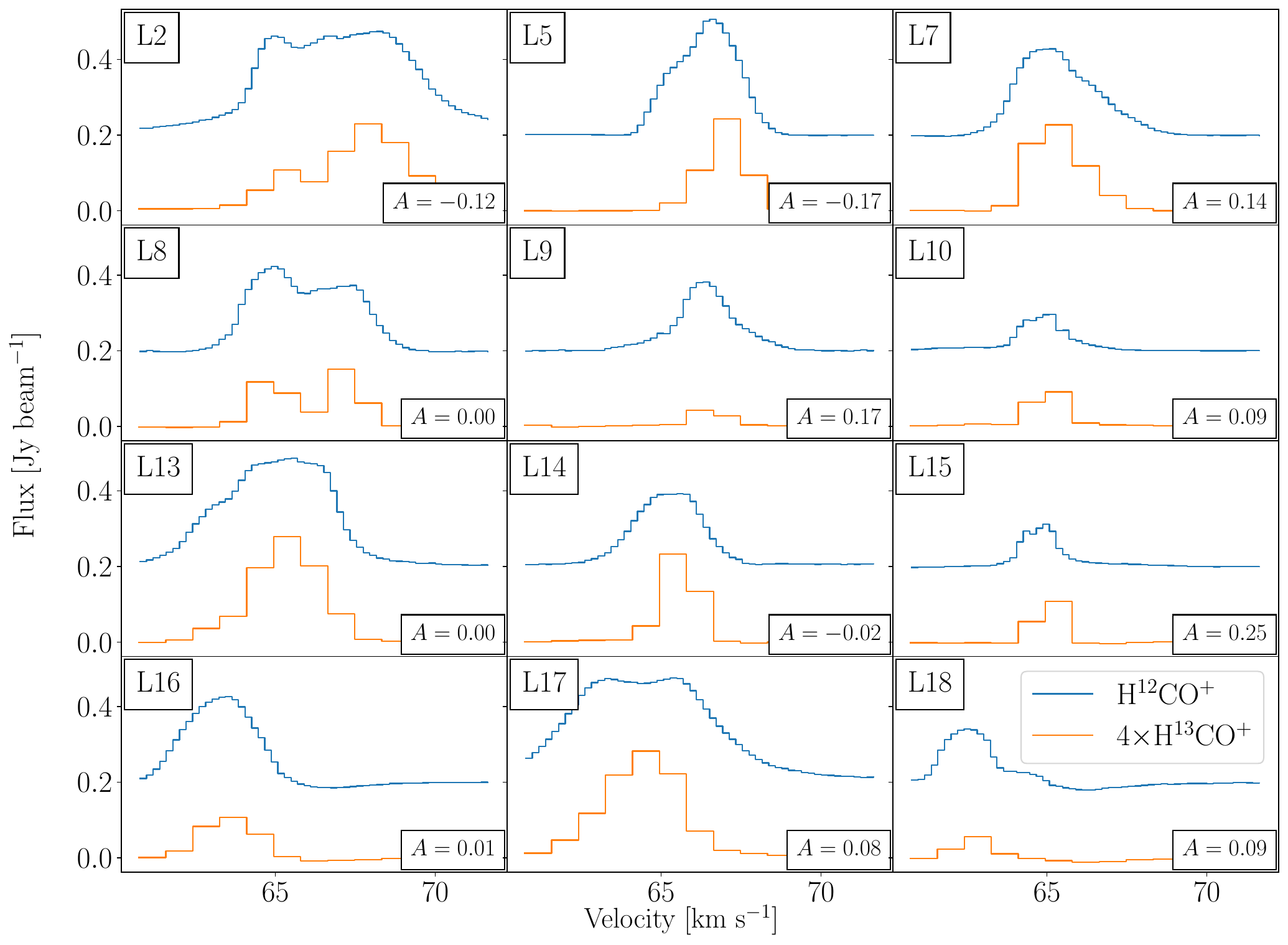}
\caption{Averaged spectral profiles of dendrogram leaves.
In each panel, the dendrogram leaf ID and asymmetry parameter ($A$) are labeled (see text for details). In all the panels, H$^{12}$CO$^{+}$ profiles are vertically offset by 0.2 Jy beam$^{-1}$, while the H$^{13}$CO$^{+}$ profiles are scaled up by a factor of 4.
} 
\label{fig5x_leafsp}
\end{figure}

To investigate the core scale gas kinematics, we have extracted the ALMA H$^{12}$CO$^{+}$ and H$^{13}$CO$^{+}$ spectral line data.
Figure~\ref{fig5x_leafsp} presents the spectral profiles of dendrogram leaves averaged over their footprint size (see Table~\ref{tab:cores}). The profiles are displayed for H$^{12}$CO$^{+}$ (optically thick) and H$^{13}$CO$^{+}$ (optically thin) emission. 
The dendrogram leaves L2 and L8 exhibit double-peaked profiles in both the H$^{12}$CO$^{+}$ and H$^{13}$CO$^{+}$ line tracers. Leaf L17 displays a double-peaked profile in H$^{12}$CO$^{+}$ only, while all other structures (i.e., L5, L7, L9, L10, L13, L14, L15, L16, and L18) exhibit single peaks.
%
The asymmetry in H$^{12}$CO$^{+}$ line profiles are indicative of significant dynamical activity \citep[e.g.,][]{Traficante2018}.
To quantitatively estimate the asymmetry in the H$^{12}$CO$^{+}$ line profiles, we derived the nondimensional asymmetry parameter ($A$) for dendrogram leaves as \citep[see more details in][]{Mardones1997}:
\begin{equation}
 A= \frac{V_{\rm thick}-V_{\rm thin}}{\Delta V_{\rm thin}},
\end{equation}
where $V_{\rm thick}$ and $V_{\rm thin}$ are the mean velocities of H$^{12}$CO$^{+}$ and H$^{13}$CO$^{+}$ line profiles, respectively. The $\Delta V_{\rm thin}$ denotes the FWHM linewidth of H$^{13}$CO$^{+}$ line profile.
We estimated the $V_{\rm thick}$, $V_{\rm thin}$, and $\Delta V_{\rm thin}$ from the Gaussian fitting of spectral profiles. The parameter $A$ for each leaf is labeled in the respective panels of their spectral profiles in Figure~\ref{fig5x_leafsp}. Most of the leaves except L2, L5, L8, L13, and L14 have positive $A$ values, indicating a blue asymmetry, which is suggestive of infall \citep[e.g.,][]{Gregersen1997,Mardones1997}.

\subsection{Distribution of Physical Parameters and spatio-kinematic analysis}
\label{str:pp}

\begin{figure*}
\includegraphics[width=\linewidth]{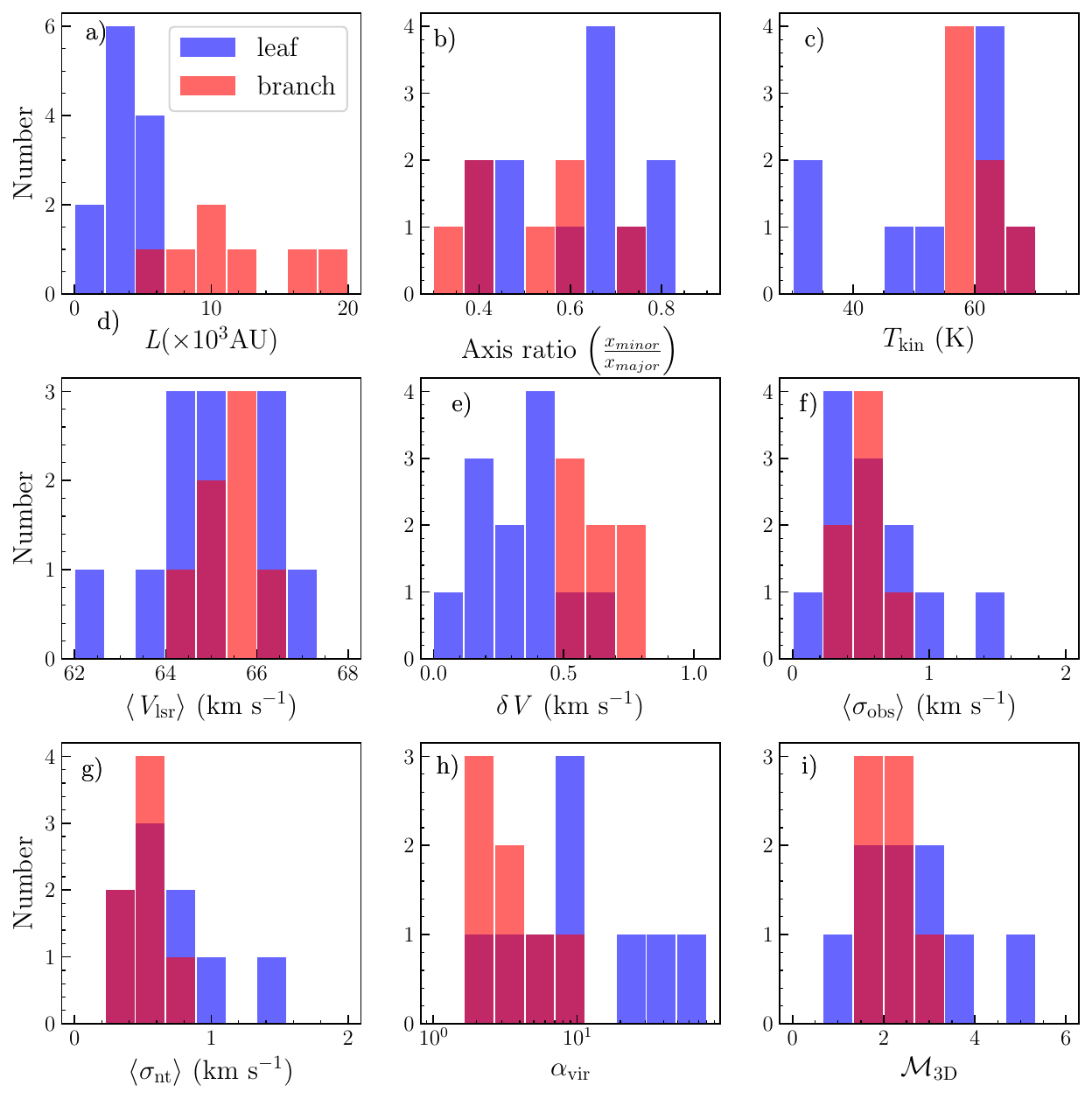}
\caption{Histograms of different properties of {\it astrodendro} identified structures. These include structure size ($\mathit{L}$ in AU), axis ratio $\left(\frac{\rm minor~axis}{\rm major~axis}\right)$, mean velocity $\left(\langle\mathit{V_{\rm lsr}}\rangle\right)$, velocity variation $\left(\delta\mathit{V_{\rm lsr}}\right)$, observed mean velocity dispersion $\left(\langle\sigma_{\rm obs}\rangle\right)$, non-thermal velocity dispersion $\left(\langle\sigma_{\rm nt}\rangle\right)$, gas kinetic temperature ($T_{\rm kin}$), virial parameter ($\alpha_{\rm vir}$), and Mach number ($\mathcal{M_{\rm 3D}}$). Histograms in blue and red correspond to leaf and branch structures, respectively.
} 
\label{fig4}
\end{figure*}

Figure~\ref{fig4} shows the histogram distribution of various physical parameters identified in Section~\ref{str:iden}. The high-resolution ALMA H$^{13}$CO$^{+}$ (4--3) data allow us to trace the structures within the physical extent of $\sim$2000 AU ($\sim$0.01 pc), revealing the core scale properties. The histogram distribution of dendrogram structure size $(L)$ is presented in Figure~\ref{fig4}a.
The branches are larger in size compared to the leaves, and consistent with the definitions of these structures (Section~\ref{dendro:analysis}).
The distribution of axis ratio $\left(\frac{\rm minor~axis}{\rm major~axis}\right)$ for identified structures is shown in Figure~\ref{fig4}b. The axis ratio for branches lies lower side than that of leaves indicating that leaves tend to be more circular (or spherical) than branches.
The gas kinematic temperature ($T_{\rm kin}$) distribution is presented in Figure~\ref{fig4}c. 
The statistical parameters related to $T_{\rm kin}$ are mentioned in Section~\ref{sec:temp}.

%

Figures~\ref{fig4}d-\ref{fig4}i display the kinematic properties of dendrogram structures. 
The weighted mean velocity of leaves and branches (see Figure~\ref{fig4}d) are found to be similar with a value of $\sim$65 km s$^{-1}$. However, the branches show narrower distribution (range = [64.53, 66.65] km s$^{-1}$) compared to leaves (range = [62.37, 67.17] km s$^{-1}$). This is possible because the fewer leaves that are separated from the central zone (i.e., leaves L10, L16, and L18; see Figure~\ref{fig3x}), are identified at lower velocities and do not have their parent branches. 
A clear separation in the $\delta V_{\rm lsr}$ distribution for leaves and branches is noticeable in Figure~\ref{fig4}e. The leaves, overall have less $\delta V_{\rm lsr}$ (range = [0.05, 0.65] km s$^{-1}$, mean= 0.34 km s$^{-1}$, median= 0.35 km s$^{-1}$) compared to branches (range = [0.56, 0.82] km s$^{-1}$, mean= 0.66 km s$^{-1}$, median= 0.67 km s$^{-1}$).

%
%

%

%
%
%

The histogram distribution of $\langle \sigma_{\mathrm{obs}} \rangle$ and its non-thermal contribution, $\langle \sigma_{\mathrm{nt}} \rangle$, for the identified structures are presented in Figures~\ref{fig4}f and~\ref{fig4}g, respectively.
The branches, overall, tend to have narrower velocity dispersion range (i.e., $\langle \sigma_{\mathrm{obs}} \rangle$; range = [0.42, 0.80] km s$^{-1}$, mean= 0.57 km s$^{-1}$, median= 0.54 km s$^{-1}$) compared to that of leaves (i.e., $\langle \sigma_{\mathrm{obs}} \rangle$; range = [0.19, 1.37] km s$^{-1}$, mean= 0.60 km s$^{-1}$, median= 0.50 km s$^{-1}$).
The similar trend is reflected in the non-thermal velocity dispersion. For branches, the $\langle \sigma_{\mathrm{nt}} \rangle$ ranges from 0.40 to 0.79 km s$^{-1}$ with mean and median values of 0.56 km s$^{-1}$ and 0.53 km s$^{-1}$, respectively, while for leaves, $\langle \sigma_{\mathrm{nt}} \rangle$ is observed in the range = [0.31, 1.37] km s$^{-1}$ with mean and median values of 0.69 km s$^{-1}$ and 0.51 km s$^{-1}$, respectively.
The $\alpha_{\rm vir}$ distribution is shown in Figure~\ref{fig4}h. Branches have less $\alpha_{\rm vir}$ (mean=3.89 and median=2.92) compared to leaves (mean=18.36 and median=10.89). 
The distribution of $\mathcal{M_{\rm 3D}}$ (see Figure~\ref{fig4}i) reveals the transonic--supersonic (1$\leq\mathcal{M_{\rm 3D}}\leq$5) nature of dendrogram structures.
The mean (median) values of $\mathcal{M_{\rm 3D}}$ for leaves and branches are 2.69 (2.39) and 2.08 (2.04), respectively.

\subsection{Line-width size ($\sigma - L$) relation}
\label{sec:vsf}

The linewidth-size ($\sigma - L$) relationship of the molecular clouds has been investigated extensively in the literature (see Section~\ref{sec:intro}).
Typically, this relationship is well described by a power law with an index ranging from 0.2 to 0.6 \citep{Larson1981,Solomon1987,Goodman1998,Heyer2004,Heyer2009,Falgarone2009,Hacar2016,Dewangan2019,Dewangan2022ionized}, and is commonly known as first of the three ``Larson's Relations''. Since the nature of turbulence and star formation is closely linked to these relationships, it is crucial to evaluate the $\sigma - L$ correlations in various environments and scales to gain a comprehensive understanding of the star-forming regions. To understand the nature of turbulence in a star-forming environment, we investigated the $\sigma_{\rm nt} - L$ relationship of regions in the immediate vicinity of W42-MME (see Figure~\ref{fig:f1}b). This is of the type--4 $\sigma - L$ relation as discussed by \citet{Goodman1998}. 
Earlier, \citet{Goodman1998} devised four types of $\sigma - L$ relations, which are the combinations of line tracers (i.e., single or multiple) and the target of interest (i.e., single or multiple molecular clouds). In this way type--1 relation is for multitracer, multicloud intercomparison, type--2 is for single-tracer, multicloud intercomparison, type--3 is for multitracer study of a single cloud, and type--4 indicates the single-tracer study of a single cloud.
Here we study this relation by two approaches, one using dispersion--size plot for dendrogram structures, and other by estimating the velocity structure function (VSF) for the entire region of study.

\subsubsection{Scaling relations from Dendrogram Structures}
\label{sec:vsf_dm}

\begin{figure}
\includegraphics[width=\linewidth]{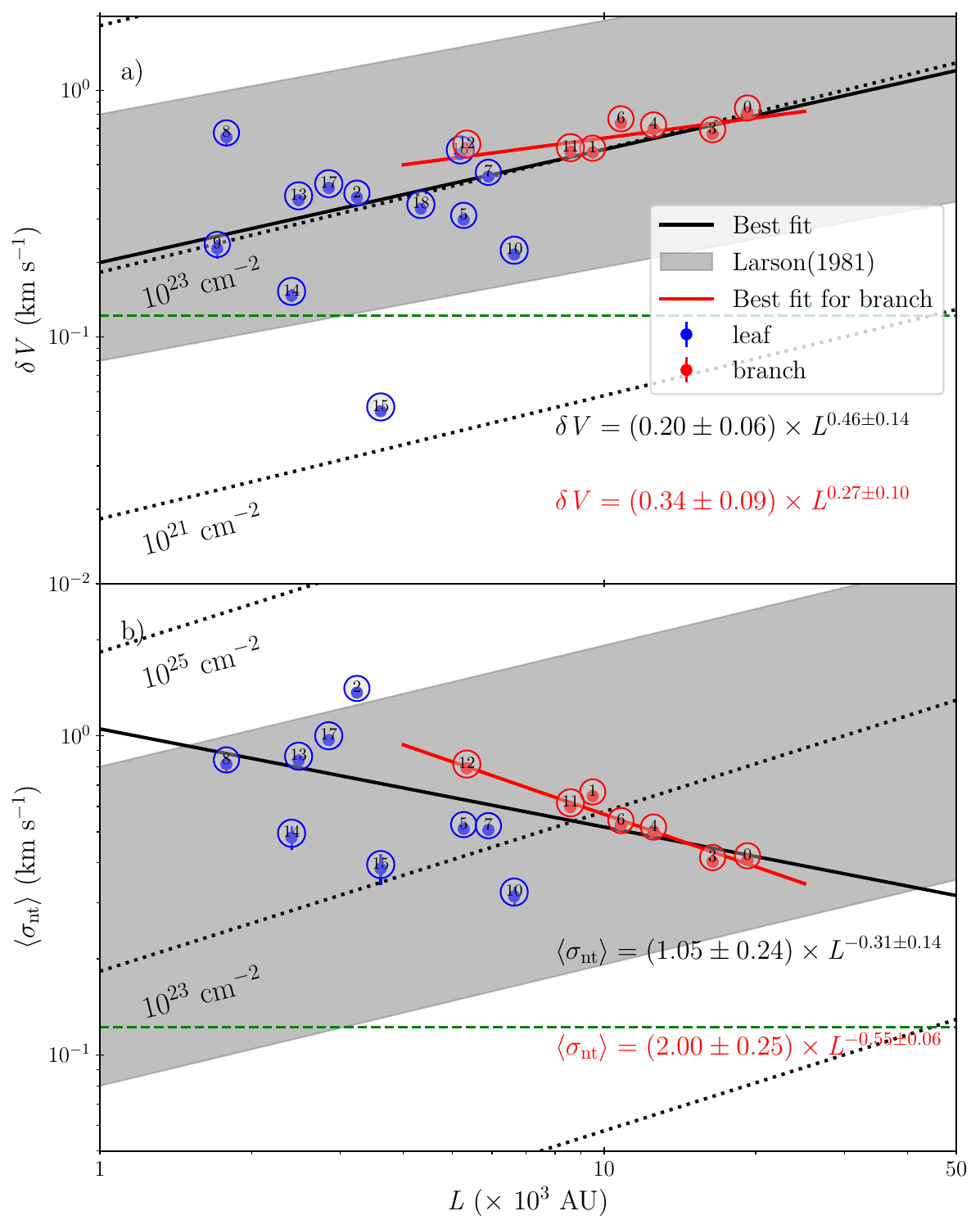}
\caption{a) $\delta\mathit{V_{\rm lsr}}$--$\mathit{L}$ plot. b) $\langle\sigma_{\rm nt}\rangle$--$\mathit{L}$ plot.
In both the panels, blue and red dots correspond to leaf and branch structures, respectively. 
We have also labeled the corresponding structure IDs.
The best fit power-law for all the dendrogram structures (solid black line) and for only branch (red solid line) are  displayed and labeled. The \citet{Larson1981} relation ($\mathit{L}\propto\sigma^{0.38}$) is shown by shaded region with the coefficient range of 0.08 to 0.8 km s$^{-1}$.
A green dashed line indicates thermal velocity dispersion of 0.129 km s$^{-1}$ at mean gas temperature of dendrogram structures (i.e., 54 K). The dotted lines denote the lines of constant column density (see labels) if non-thermal motions are driven by the self gravity.
} 
\label{fig5}
\end{figure}

In the analysis below, we explore two scaling relationships concerning the dendrogram density structures (i.e., leaves and branches) identified in W42 region. 
Two parameters, $\sigma$ and $\delta V_{\mathrm{lsr}}$ are used to measure internal gas motions \citep[e.g.,][]{Storm2014,Liu2022}.
The parameter $\sigma$ indicates the internal gas motion in the direction of an individual structure (line-of-sight velocity dispersion or linewidth), while $\delta V_{\mathrm{lsr}}$ traces the gas motion across different directions within the structure (plane-of-sky velocity dispersion).
At the cloud scale (>10 pc), the choice of these parameters is insignificant as these two parameters are similar and commonly trace the internal gas motion of the cloud. However, at smaller scales (<0.1 pc), local effects become more important and gas properties become distinguishable.
Therefore, in particular, we are interested in the scaling relationships of $\delta\mathit{V_{\rm lsr}}-L$ and $\sigma_{\rm nt} - L$.
Figure~\ref{fig5}a presents the $\delta\mathit{V_{\rm lsr}}-L$ plot for the identified dendrogram structures. 
The best fit power-law has the form $\delta\mathit{V_{\rm lsr}}=(0.20\pm0.06)\times L^{0.46\pm0.14}$, which closely follows the generalized Larson's law with power-law scaling exponent of $\sim$0.5 at a constant column density of $\sim$10$^{23}$ cm$^{-2}$ \citep[e.g.,][]{Heyer2004}.
We have also marked the lines of constant column density in Figure~\ref{fig5} following the equation, $\sigma=\sqrt{2G\Sigma L}$, where $\Sigma$ is the column density, $G$ is the universal gravitational constant, and $L$ is the structure size \citep[see][for details]{Ballesteros-Paredes2011}.
This equation is valid if the collapsing scenario is applicable to all scales within the molecular cloud and the non-thermal motions are of gravitational origin. However, such equation is also applicable for the case of virial equilibrium, with a difference by a factor of $\sqrt{2}$ less \citep[see more details in][]{Ballesteros-Paredes2011}. 
The power-law exponent in the $\delta\mathit{V_{\rm lsr}}-L$ relation for branch structures is found to be shallower with the value of 0.27$\pm$0.10 compared to the global trend.
We have marked the footprint of Larson's law with the scaling coefficient ranging from 0.08 to 0.8 by shaded area in Figure~\ref{fig5}.

We further estimated the Pearson Correlation Coefficient (PCC) which is a measure of linear correlation between two sets of data. 
The PCC correlation coefficient $r$ indicates the strength of the linear relationship, while the significance is measured by the probability value $p$.
It is to be noted that the $r$ ranges from $-1$ to $+1$ signifying the strong anti-correlation ($r=-1$), no correlation ($r=0$), and strong positive correlation ($r=+1$), respectively.
The smaller $p$ value indicates more significant relationship. In general $p<$0.05 signifies statistically significant correlation.
For $\delta\mathit{V_{\rm lsr}}-L$ plot, we obtained the $r$--value and $p$--value of 0.69, and 2.92$\times$10$^{-3}$, respectively. This indicates the positive correlation between $\delta V_{\mathrm{lsr}}$ and $(L)$ of dendrogram structures.

Figure~\ref{fig5}b shows the $\sigma_{\rm nt} - L$ plot for the identified structures. This plot overall shows negative trend with scaling exponents of $-$0.31$\pm$0.14, which is obtained from the best fit power-law over all the identified structures. 
With $p=0.056$, the correlation coefficient $r=-0.4$ between $\sigma_{\rm nt}$ and $L$ is marginally statistically significant.
In $\sigma_{\rm nt} - L$ plot, a steeper negative trend with a power-law exponent of $-$0.55$\pm$0.06 is found for the branch structures only. The PCC $p-$value of 2.04$\times$10$^{-3}$ with $r=-0.93$ suggests its statistical significance.
The possible origin of these outcomes are discussed in Section~\ref{sec:dis}.

\subsubsection{Velocity Structure Function}
\label{sec:vsf_reg}

An important statistical description of interstellar turbulence or the gas dynamics can be inferred by the generalized VSF \citep[e.g.,][]{Miesch1994,Heyer2004,Chira2019}
\begin{equation}
S_{p}(l) = \langle \left|(\varv(r) - \varv(r+l)\right)|^p\rangle,
\end{equation}
where, $l$ is known as lag (i.e., the spatial displacement between two points within a three dimensional volume), $p$ is the order of VSF, and the average is taken out for the entire volume of the gas. The VSF is described as a power-law over a finite spatial range and often reframed as a power-law expression by taking $p^{th}$ root; i.e., $S_{p}(l)^{1/p} = \varv_{o}~l^{\gamma}$, where $\varv_{o}$ is a scaling coefficient and $\gamma$ is a scaling exponent \citep{Heyer2004}.
In our analysis, we calculated the reframed 2$^{nd}$ order structure function in velocity $\delta{v}=S_{2}(l)^{1/2}$ \citep[e.g.,][]{Hacar2016,Dewangan2019} using H$^{13}$CO$^{+}$(4--3) data of entire region of our study (see Figure~\ref{fig:f1}b). 
In our calculations, we adopted only those data points which have peak intensities greater than 0.01 Jy beam$^{-1}$ and have velocity dispersion (FWHM) $<$ 2.5 km s$^{-1}$. Applying these conditions, the total number of data points were 7566 (i.e., $\sim$11\% of available points). We then estimated $S_{2}(l)^{1/2}$ using moment-1 map of H$^{13}$CO$^{+}$ and the total range of angular separations (i.e., 0$''$.05--19$''$.2) were divided into sub-ranges (or lags) of 0$''$.13 (or $\sim$500 AU in spatial scale).

\begin{figure}
\includegraphics[width=\linewidth]{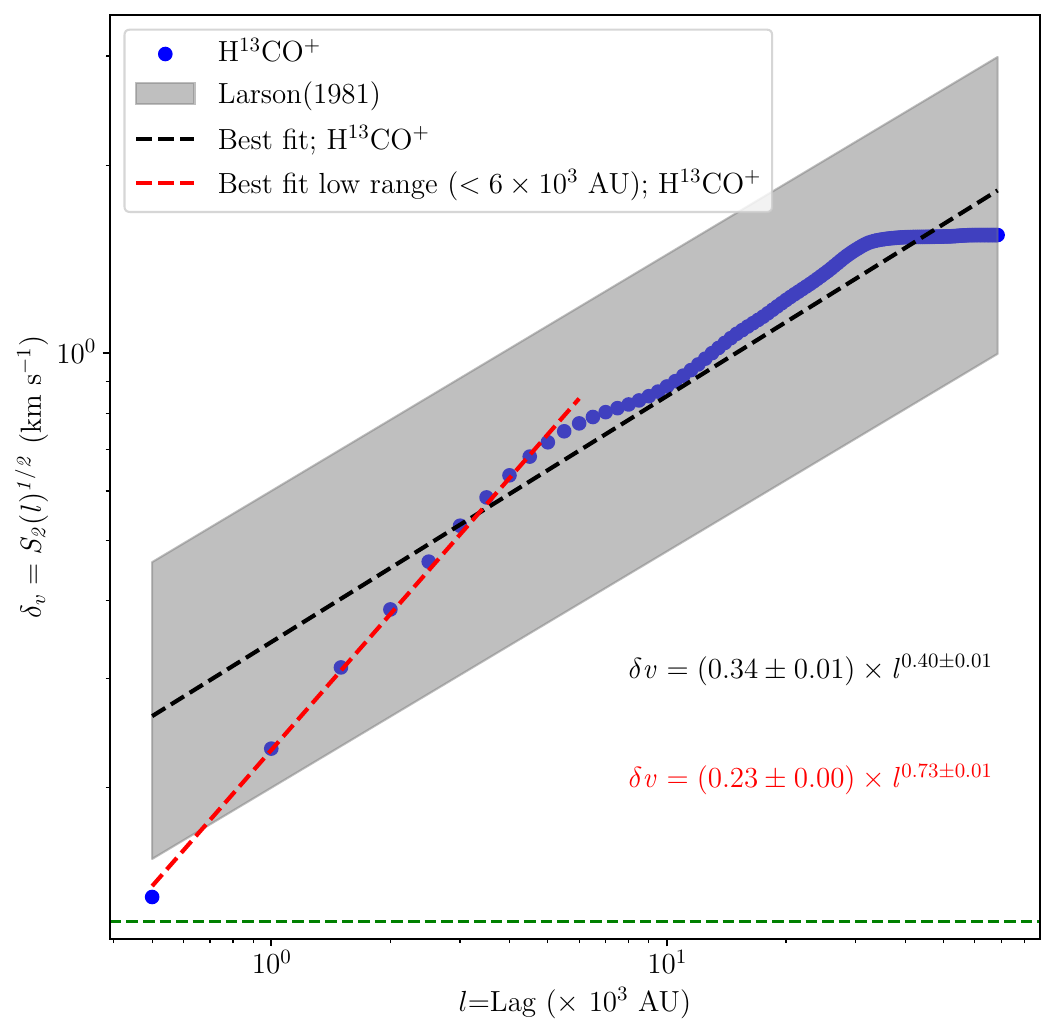}
\caption{Structure function in velocity ($\mathit{S_{2}(l)^{1/2}}$) as a function of lag $(l)$.
The lag is sampled at the steps of 500 AU. A green dashed line indicates $\sigma_{\rm th}$ of 0.129 km s$^{-1}$ at mean gas temperature of 54 K.
} 
\label{fig5x}
\end{figure}

Figure~\ref{fig5x} shows the $S_{2}(l)^{1/2}-l$ plot for H$^{13}$CO$^{+}$ (blue points) data. 
We have estimated the best fit power-law form of H$^{13}$CO$^{+}$ $S_{2}(l)^{1/2}$, which is obtained as $S_{2}(l)^{1/2}=(0.34\pm0.01)\times l^{0.40\pm0.01}$. Here we also note that there is no significant deviation in the $S_{2}(l)^{1/2}$ for H$^{13}$CO$^{+}$ and H$^{12}$CO$^{+}$ data (not shown here). In general, the $S_{2}(l)^{1/2}$ shows a linear trend in the spatial range within $\sim$6000 AU. The form of this linear trend is obtained as $S_{2}(l)^{1/2}=0.23\times l^{0.73\pm0.01}$, and is more steeper than the global trend with the form of $S_{2}(l)^{1/2}\propto l^{0.40\pm0.01}$. However, for larger separations ($\gtrsim$6000 AU), the VSF rises slightly, and shows oscillatory behavior. This form of VSF is seen previously by other authors as well \citep[e.g.,][]{Hacar2016,Dewangan2019,Dewangan2022ionized}.

\begin{figure}
\begin{subfigure}{1\linewidth}
\includegraphics[width=\textwidth]{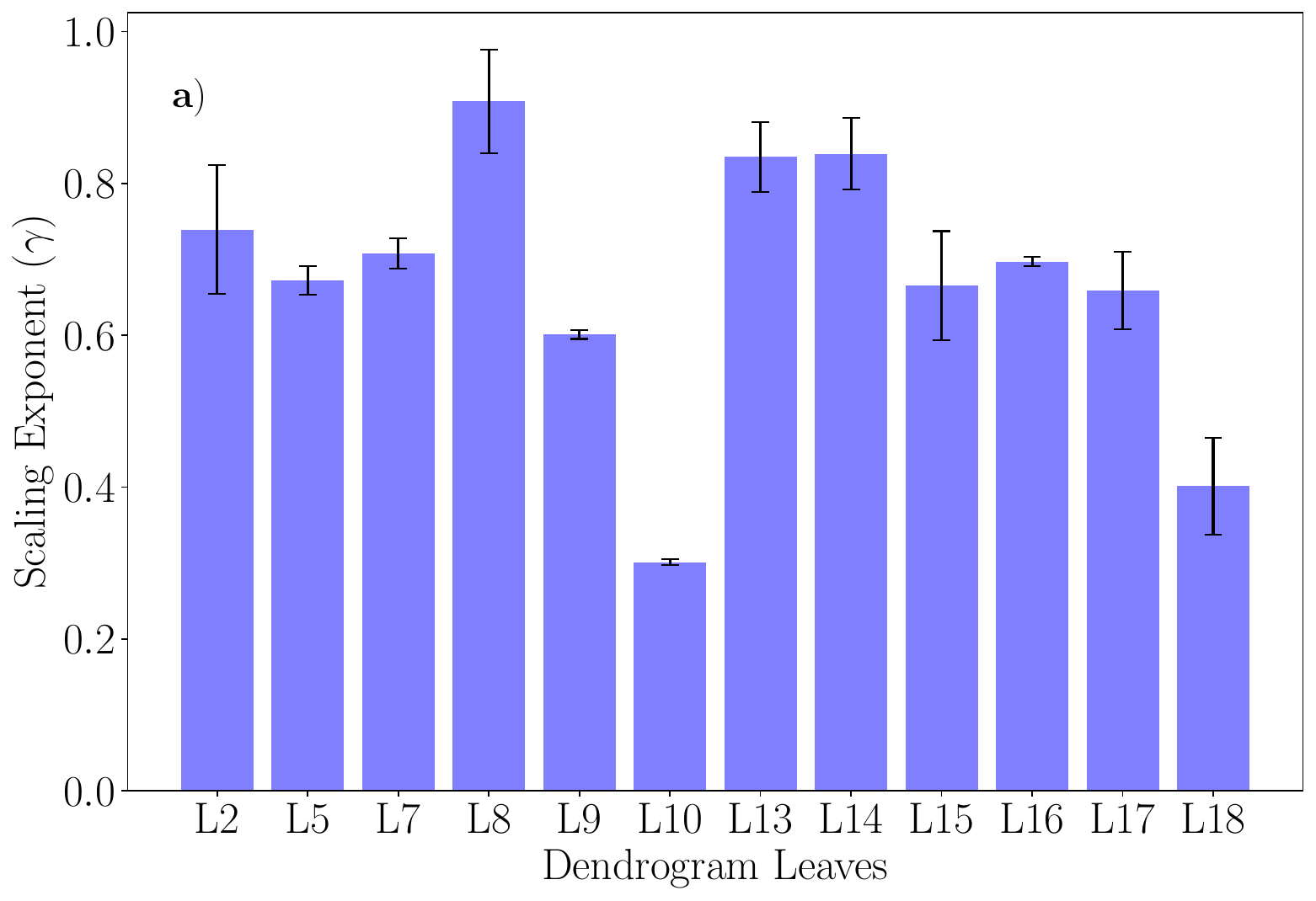}
\end{subfigure}

\begin{subfigure}{1\linewidth}
\includegraphics[width=\textwidth]{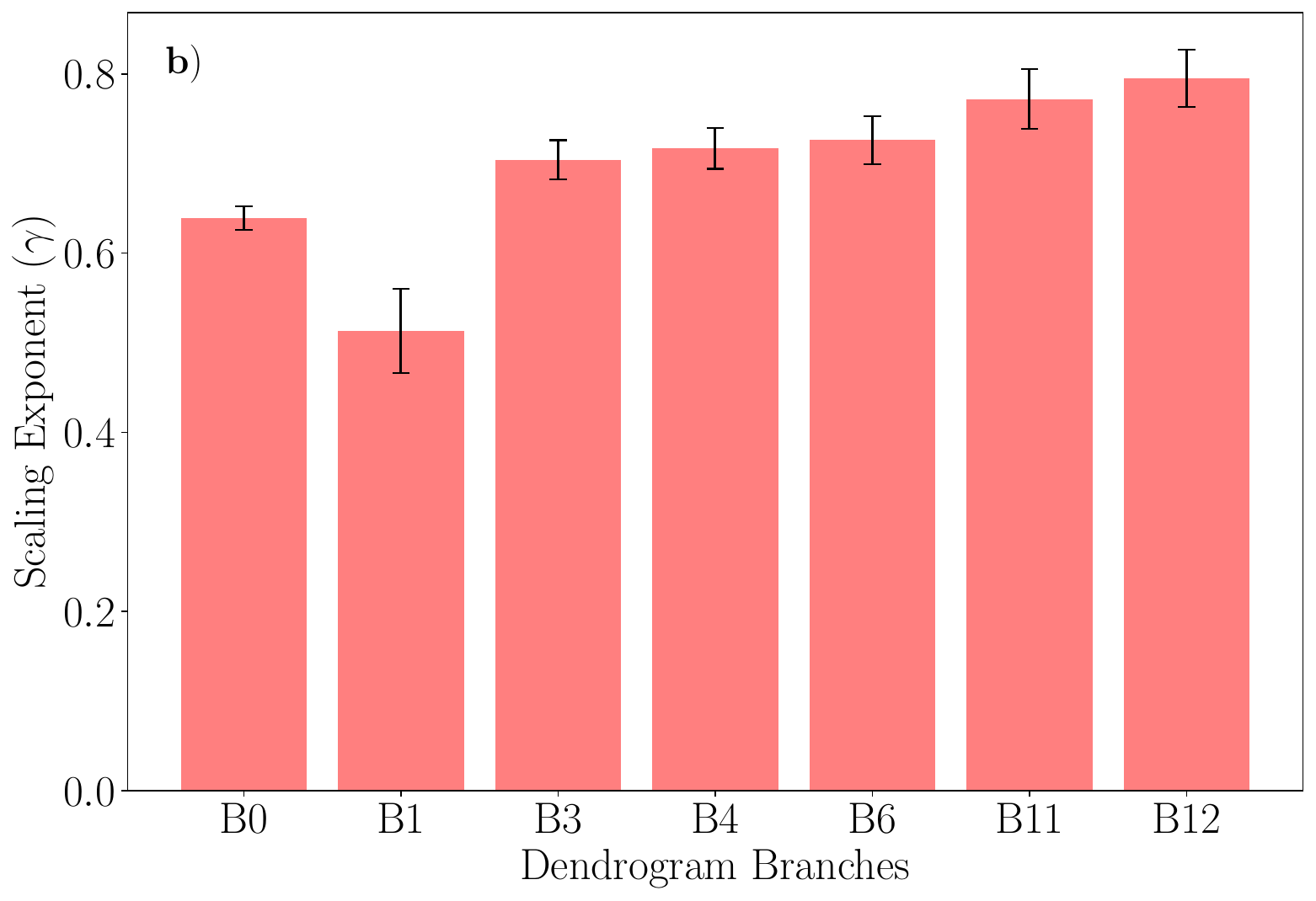}
\end{subfigure}

\begin{subfigure}{1\linewidth}
\includegraphics[width=\textwidth]{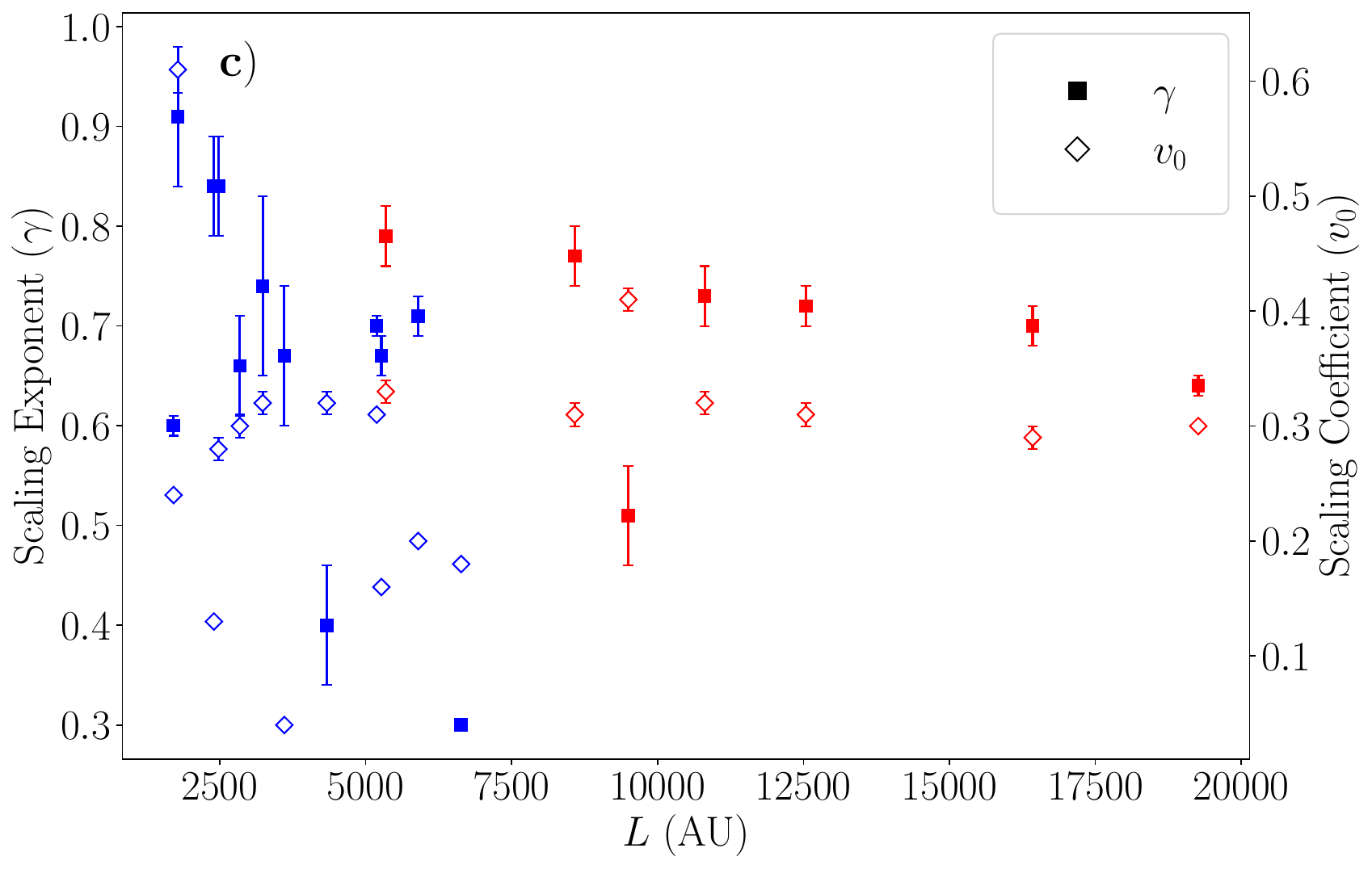}
\end{subfigure}

\caption{Distribution of scaling exponent ($\gamma$) defined in the VSF of second order $S_2(l)^{1/2}=\delta v= v_{0}l^{\gamma}$ for a) dendrogram leaves and b) branches. c) Variation of $\gamma$ and $\varv_{0}$ with respect to the structure size. Blue and red colors indicate the leaves and branches, respectively.
The parameters $\gamma$ and $\varv_{0}$ are obtained from the power-law best fit for the $S_2(l)^{1/2}$ of each dendrogram structure. 
The errors denote 1 sigma standard errors on the fitted parameters.
} 
\label{fig8}
\end{figure}

We have further investigated the variation of scaling coefficient ($\varv_{o}$) and scaling exponent ($\gamma$) for each dendrogram structure by computing the H$^{13}$CO$^{+}$ $S_{2}(l)^{1/2}$ in Figure~\ref{fig8}. 
The histograms of $\gamma$ for leaves and branches are shown in Figures~\ref{fig8}a and \ref{fig8}b, respectively.
Here we note that the $\gamma$ is related to a linear regime in log-log plots of $S_{2}(l)^{1/2}$ of each structure (for comparison, see Figure~\ref{fig5x}).
To perform such an analysis, we first masked out the area that does not include our region of interest (i.e., each dendrogram structure) and then computed the VSF.
Due to the small structure sizes, here the VSF analysis is exempted from the conditions defined earlier (i.e., imposing cut-off in peak intensity and velocity dispersion). 
The total number of data points range from 62 to 943 for leaves, while for branches the data points were in the range of 675--5947.
Figure~\ref{fig8}c displays the distribution of $\gamma$ and $\varv_{o}$ as a function of structure size. The distribution of $\gamma$ is scattered for leaves, while it shows a negative trend for the branches. The distribution of $\varv_{o}$ is nearly constant for the branches.


%

\subsubsection{Mass--size and density--size relations}

\begin{figure}
\includegraphics[width=\linewidth]{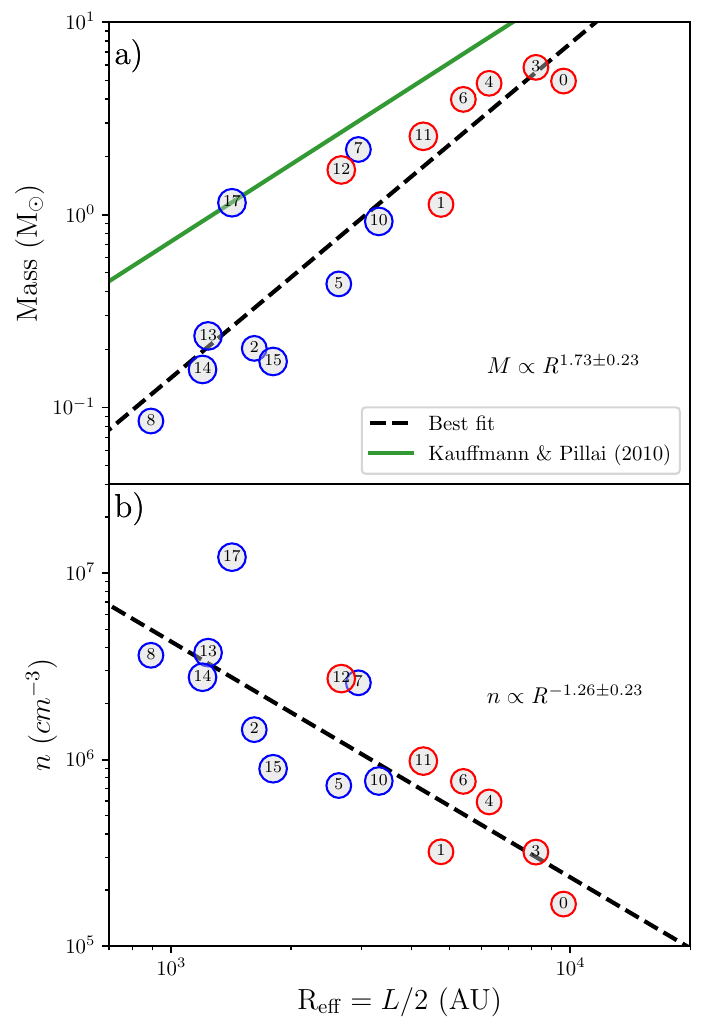}
\caption{a) Mass--size relation and b) density--size relation of dendrogram structures. Blue (red) circles with labels denote the leaf (branch) structures. A green line indicates the mass--size law of $m(r)> 870~M_{\odot} (r/pc)^{1.33}$ for massive star forming clumps proposed by \citet{Kauffmann2010}, while dashed line in each panel represents the best fit. The best fit equation is also labeled in each panel.
} 
\label{fig5x_ms}
\end{figure}

To understand the nature of detected dendrogram structures, we computed the mass--size $(M-R)$ and density--size $(n-R)$ relations for the dendrogram structures in Figure~\ref{fig5x_ms}. 
These relations can be derived from one another, implying that the column density of structures with various masses, sizes, and evolutionary states is constant. Collectively, these relations are known as Larson's third relation.
The $M-R$ relation has a best fit power-law with a form of $M\propto R_{\rm eff}^{1.73\pm0.23}$, where $R_{\rm eff}$ is the effective radius of dendrogram structures and defined earlier. The obtained PCC $r$--value and $p$--value of $M-R$ plot are 0.92, and 5.4$\times$10$^{-7}$, respectively. This indicates the strong positive correlation between mass and radius of dendrogram structures, which is consistent with previous studies \citep[e.g.,][]{Kauffmann2010ApJ1,Kauffmann2010ApJ2,Kauffmann2010,Ballesteros-Paredes2019}.
In the case of nearby molecular clouds, the $M-R$ power-law index is close to 2 \citep[e.g.,][and references therein]{Lombardi2010,Lada2020,Cahlon2023arXiv230814794C}. For distant clouds spread over the entire Galaxy this index is found to be $>$2 \citep[e.g.,][]{Miville2017, Traficante2018larson}. However, \citet{Ballesteros-Paredes2019} suggest that the larger power-law index could result from the superposition of line-of-sight dust emission. Additionally, they also pointed out that the $M-R$ relation is dependent on how the cloud structure is defined, e.g., the column density cutoff leads to the power-law index of 2, while the volume density definitions results in the index value of 3 \citep[see also][]{Cahlon2023arXiv230814794C}.

Earlier, \citet{Kauffmann2010} proposed an empirical mass--size threshold for MSF which is based on the clouds with and without MSF. These authors showed that density structures with $m(r)> 870~M_{\odot} (r/pc)^{1.33}$ are prone to form massive stars. In our analysis, we noticed that only leaf L17 lies within this relation. Infact, leaf L17 was earlier proposed and confirmed as young massive protostar by \citet{Dewangan2015b,Dewangan2022} \citep[see also][]{Buizer2022}.
In a similar way, a plot of mass vs. surface density ($\Sigma=\frac{M}{\pi R_{\rm eff}^{2}}$; not shown here) displays a scatter with only leaf L17 statisfying the condition of MSF with $\Sigma>1$ g cm$^{-2}$ \citep[e.g.,][]{Urquhart2014,Saha2022}. 

Figure~\ref{fig5x_ms}b shows the $n-R$ relation for dendrogram structures. The PCC $r$--value and $p$--values are $-$0.5, and 0.05, respectively implying the marginally statistically significant anti-correlation between density and size. 
The power-law form of density--size relation is observed as $n\propto R_{\rm eff}^{-1.26\pm0.23}$. Here we note a caveat that some of the anti-correlation in the  $n-R$ relation is there by construction, given that both the axes depend on the value of $R_{\rm eff}$ (i.e., $L=2 \times R_{\rm eff}$ and $n_{\rm H_2} = \frac{M_{\rm s}}{(\frac{4}{3}\pi R_{\rm eff}^3) (2.8 ~m_{\rm H})}$ ;see Section~\ref{sec:mass}).

%

\subsection{ALMA 865 $\mu$m dust continuum cores}
\begin{figure}
\includegraphics[width=\linewidth]{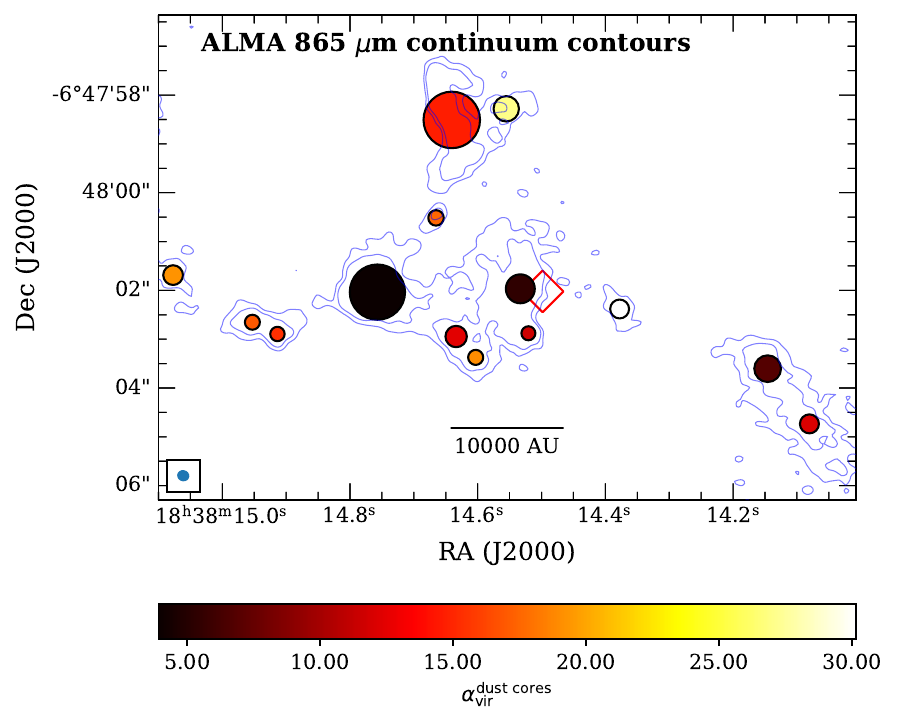}
\caption{ALMA 865 $\mu$m continuum emission contours (levels=[0.4, 1] mJy beam$^{-1}$) overlaid with the positions of dust cores. 
The size of dots is proportional to the footprint area of dust cores identified in ALMA 865 $\mu$m continuum emission map.
Color scale signifies the virial parameter of dust cores.
The beam size and a scale bar is shown in the image.
} 
\label{fig5x_vir}
\end{figure}

In Section~\ref{str:iden}, we extracted dendrogram leaf structures from the ALMA H$^{13}$CO$^{+}$ emission and estimated their physical properties, including mass, virial mass, and virial parameter. We noticed that larger values of virial parameters can arise due to the lack of significant dust emission within structures where H$^{13}$CO$^{+}$ emission is observed, as shown in Table~\ref{tab:cores}. 
To address this issue, we identified dust cores (or leaf structures in dust continuum map) using the dendrogram methodology on the ALMA 865 $\mu$m continuum map and estimated the aforementioned physical parameters. In order to identify the cores, the parameters ``min\_value'' and ``min\_delta'' were set to be 0.5 mJy beam$^{-1}$ and 0.2 mJy beam$^{-1}$, respectively. The parameter ``min\_npix'' was chosen such that the cores contains at least 2 beams of the ALMA 865 $\mu$m map (see Section~\ref{str:iden}). The analysis resulted in the identification of 14 cores. These dust cores are found to be spatially coexisted with the leaf structures identified in H$^{13}$CO$^{+}$ emission (for comparison see Figures~\ref{fig3x} and ~\ref{fig5x_vir}). 
The size of these dust cores ranges from 1700--6850 AU with mean and median values of 3000 and 2340 AU. 
The observed mass of dust cores varies from 0.04 to 1.64 $M_{\odot}$, and is relatively lower than the mass of H$^{13}$CO$^{+}$ dendrogram leaves (i.e., 0.08--2.1 $M_{\odot}$; see Table~\ref{tab:cores}).
However, the observed mass of these small scale structures is possible because of the missed continuum flux at high resolution ALMA observations, as the interferometric observations are sensitive to strong and compact sources but miss more diffuse and extended emission. Based on this statement the statistics could be slightly biased.
Moreover, on scales of $\sim$1000 AU, gas-to-dust mass ratios could differ significantly from the canonical value of 100 used for mass calculations on larger scales \citep{Weingartner2001}. 
Thus, the low values of mass for these structures (including L17, hosting MYSO) is alone insufficient to discuss their star-forming nature. 
Also, earlier study by \cite{Dewangan2022} suggest that the MYSO W42-MME continues to gain mass from its surroundings, which may be true for other dense structures.
Figure~\ref{fig5x_vir} presents an overlay of the identified core positions by dots on the ALMA 865 $\mu$m map displayed with continuum contours. The dot size is proportional to the structure's footprint area, and the color scale represents the virial parameter information.
To estimate the virial mass, we used a mean gas temperature of 54 K, and the mean velocity dispersion for each core was derived using the $M_{2}$ map of the ALMA H$^{13}$CO$^{+}$ emission (see Section~\ref{str:phyparam}). The resulting virial parameter values for the dust cores ranged from $\sim$4 to 30, with mean and median values of 15.14 and 14.95, respectively. These values indicate that the dense cores in the W42 region are in overvirial state.

\section{Discussion}
\label{sec:dis}

To examine the physical surroundings of the young massive protostar W42-MME, which is enclosed in a dusty cocoon/envelope ($\sim$9000 AU) and influenced by the radiative feedback from a nearby massive O5.5-type star \citep{Dewangan2022}, the investigation of the dense gas kinematics is crucial. Our target site is important since it experiences different environmental conditions compared to young massive protostars that do not belong to the feedback-affected regions by existing nearby massive star \citep{Liu2022_atomsV, Saha2022}. It is worth mentioning that our target site is a part of large filament-hub \citep[extent $\sim$1--2 pc; see Figure~14 in][]{Dewangan2022}. These HFSs are widely recognized among the research community as the initial stages of MSF \citep[e.g.,][and references therein]{Motte2018,Kumar2020,Bhadari2022,Maity2022,Dewangan2023JApA,Dewangan2023}. Therefore, the study of hierarchical structures of clouds, including clumps and cores, is crucial in understanding how dense cores form within the cloud and the underlying factors involved in this process.

\subsection{Dynamical state of dense structures}

The accurate study of the dynamics of clouds and dense structures in space relies on understanding the nature of gas motion. 
The leaf structures identified in the dendrogram analysis of ALMA H$^{13}$CO$^{+}$ data are found to have relatively higher line-of-sight velocity dispersion compared to the branches (see Figure~\ref{fig4}). Additionally, all of the identified structures exhibit transonic--supersonic gas motions (i.e., 1<$\mathcal{M}$<5).
These observations are also reflected in the negative trend in $\sigma_{\rm nt}-L$ plot (see Figure~\ref{fig5}). Such a negative trend in the $\sigma_{\rm nt}-L$ relation is not commonly observed in star-forming clouds. 
However, some studies show either shallower power-law slope or no correlation as well \citep{Caselli2002,Traficante2018larson}.
Widely studied Larson's first law, on the other hand, suggests a different trend, where large fragments are expected to have higher dispersion compared to smaller ones. 
The Larson's relation is found to be consistent at larger scales \citep[$>$ 0.1 pc, e.g.,][]{Falgarone2009,Ballesteros-Paredes2011}, but at a smaller scales the studies are limited \citep[e.g.,][]{Li2020,Liu2022_atomsV,Saha2022,Li2023arXiv230401718L}.
In this context, our observed $\sigma_{\rm nt}-L$ relation appear to support the theoretical framework proposed by \citet{Ballesteros-Paredes2018}, which emphasizes that the generalized Larson's relation can have a gravitational origin. Furthermore, for varying column densities, the $\sigma_{\rm nt}-L$ relation need not necessarily follow Larson's law.
The authors suggest that as the cores collapse, their sizes become smaller, while the column densities and velocity dispersion become larger. This results in an reverse/oblique trend compared to Larson's relation \citep[see Figure~2 in][]{Ballesteros-Paredes2018}, which can be explained by the consequences of the hierarchical and chaotic collapse scenario.
This suggests the emergence of local collapse centers within the entire cloud as it itself undergoes collapse. In such case, the gravitational energy converts to the kinetic enery and developes the virial-like relation \citep[see Figure 2 in][]{Ballesteros-Paredes2018}.
We observed the infall gas signature in most of the leaf structures (see Figure~\ref{fig5x_leafsp}). Additionally, the identified structures exhibit overvirial ($\alpha_{\rm vir}>2$) states, which, as found, are similar to the time evolution of collapsing cores \citep{Ballesteros-Paredes2018}.
These results, present an ambiguous picture if we compare our results with the literature provoking that branches (or clumps) tend to be more turbulent than the leaves (or cores). 
However, some studies support our observations and advocate the overviriality of cores \citep[][and references therein]{Ballesteros-Paredes2018,Traficante2018larson,Singh_ayushi2021ApJ,Galeano2022}.
According to \citet{Ballesteros-Paredes2018}, the excess overvirial states is because of the assumption that the gravitational energy is estimated by the energy of isolated homogeneous sphere. They also stress out that such excess disappears when the gravitational energy is correctly estimated from the actual spatial mass distribution. Therefore, the Larson's $\sigma_{\rm nt}-L$ relation does not hold when the column densities spanning a large dynamic range are considered.
In general, the deviation from the Larson's scaling can be the signature of gravity-driven chaotic gas motions as predicted by global hierarchical collapse (GHC) model \citep{Ballesteros-Paredes2011,Semadeni2019}. 
Our observed scaling relations suggest that the gas motion in the vicinity of W42-MME is transonic-supersonic. 
The supersonic nature of molecular clouds can arise because of random turbulent motions, gravity-driven chaotic motions, or even ordered motions like localized rotation and outflows, or large-scale directional gas flows \citep[e.g.,][]{Li2016MNRAS,Traficante2018,Traficante2018larson}. 
In general, the point sources are not promisingly evident in our target region (see Figure~\ref{fig3x}c), however the dense gas profiles convey the signatures of infall/outflow activities (see Figure~\ref{fig5x_leafsp}). These processes can also give rise to high-dispersion values at the smaller scales.

In another approach, the $\delta V_{\mathrm{lsr}}-L$ relation for all the dendrogram structures and $S_{2}(l)^{1/2}$ for entire region of study, reveal a positive trend and closely follow the generalized Larson's relation \citep[e.g.,][]{Heyer2004} with a nearly constant column density of $\sim$10$^{23}$ cm$^{-2}$ and positive scaling exponents of 0.46$\pm$0.14 and 0.30$\pm$0.01, respectively. $S_{2}(l)^{1/2}$, however, tend to be more linear in the regime of $<$6000 AU with steeper scaling exponent of 0.68$\pm$0.02. 
This power-law dependence may indicate general behaviour of velocity field of the gas in the viscinity of W42-MME. It may be possible that the supersonic vortices created by gas motion primarily dominate at the physical scales of leaves \citep[e.g.,][]{Dewangan2019}.
The power-break in $S_{2}(l)^{1/2}$ at some length scales has no clear boundary, but our results suggest that the spectral break can occur at around $\sim$6000-8000 AU scale. This finding requires verification through more studies conducted at the physical scales of cores (i.e., $\lesssim$0.05 pc).
The spatial distance of $\sim$6000 AU, interestingly, is closely matched with the maximum structure size of dendrogram leaves (i.e., $\sim$6600 AU). 
The variation in $\gamma$ defined in the VSF of second order $S_2(l)^{1/2}=\delta v= v_{0}l^{\gamma}$ for dendrogram leaves is found to be larger compared to that of branches (see Figure~\ref{fig8}). 
This suggests that the local gas motions driven by embedded sources and associated activities (i.e., inflow/outflow) primarily contributes in shaping the VSF power-law over time evolution \citep[e.g.,][see Figure 1 therein]{Chira2019}. For larger fragments of clouds (e.g., clumps), the effect seems to be averaged out and hence the structures tend to show a less scatter in the $\gamma$. 


Here we note that both $S_2(l)^{1/2}$ and $\delta V$ (indicating the standard deviation of mean velocities within the spatial structures) closely follow the similar statistics and infer the  information of gas motion across the sky plane. Hence they appear to show similar positive trends with $L$. 
However, $\sigma$, which represents the weighted mean of standard deviations of line profiles at each pixel within the structure, provides information about the gas motions along the line-of-sight. At larger spatial scales, all these parameters can be equivalent as the gas kinematic properties will be averaged out, but at smaller scales where we expect localized gas motion, they may indicate different physical phenomena.


Overall, our observations support theoretical insights from studies that stress the non-universality of Larson's relations \citep{Ballesteros-Paredes2018}. In contrast to this, several previous studies support that turbulence plays important role at large scales while gravity primarily works at smaller scales.
We, however, suggest that the roles of turbulence and gravity cannot be differentiated by non-thermal gas motion alone at the spatial scales of sub-parsec regime, primarily in the environment where the feedback processes are dominant \citep[e.g.,][]{Traficante2018}. Earlier, \citet{Goodman2009} suggested that, apart from the well-cited role of turbulence at large scales, self-gravity plays a significant role in all possible scales of cloud fragments. Similarly, \citet{Ballesteros-Paredes2011} and their subsequent studies \citep{Ballesteros-Paredes2018,Ballesteros-Paredes2019} emphasize the role of turbulence at smaller core scales. These studies suggest that velocity dispersion in molecular clouds do not entirely depend on spatial scales, but also include surface gas densities.
Hence, more observational studies of star-forming regions at sub-parsec scales are required to clearly understand the roles of gravity and turbulence (i.e., driving factors of cloud dynamics).


\subsection{Star formation Scenario}

The multi-wavelength study at larger scale of W42 region by \citet{Dewangan2015a} shows that the massive O5.5 star and W42-MME form at a $\sim$3$\times$3 pc$^{2}$ junction of several converging filaments ($\sim$1--3 pc). 
This junction itself is situated at the waist of a bipolar H{\sc ii} region.
This picture emphasize the role of filaments in MSF as discussed by \citet{Motte2018} in their evolutionary scheme of MSF. The HFSs are widely popular and thought to provide a suitable environment for the formation of massive stars \citep[e.g.,][]{Kumar2020}. However, within a hub one can expect the interplay of different physical processes driven by gravity, turbulence, and magnetic fields. In our site, the young massive protostar W42-MME resides in close proximity to the O5.5 star \citep{Dewangan2022}. The massive O5.5 star drives enough mechanical energy in the form of its pressure components \citep[i.e., from H{\sc ii} region, radiation, wind, and by self-gravity; see][for detailed calculations]{Dewangan2015a} at the vicinity of W42-MME, thus can overall increase the turbulent properties of gas.

A filamentary morphology of coherent velocity structure (61--66 km s$^{-1}$) of size $\sim$0.3 pc spanning east-west direction is observed in H$^{13}$CO$^{+}$ data, containing the dendrogram leaves L18, L16, L7, L17 (including L13, L14), L15, and L10 (see Figures~\ref{fig3x}b and \ref{fig3}b). It is interesting to note that the spatial separation between the leaf pairs of L17--L7 ($\sim$3$''$.4) and L17--L15 ($\sim$2$''$.4) is more or less similar. Similarly, it is true for leaf pairs of L17--L16 ($\sim$6$''$.1) and L17--L10 ($\sim$6$''$.6). This may hint at the filament fragmentaion by the combined effect of gravity and turbulence \citep{Inutsuka1992,Nakamura1993,Inutsuka1997}.
Earlier \citet{Dewangan2022} pointed out the presence of at least 5 continuum peaks within a dusty envelope (or branch B11 in our analysis) that hosts W42-MME. We note that only one continuum peak \citep[see peak B in Figure~4 in][also see Figure~\ref{fig5x_vir}]{Dewangan2022} falls within the ALMA beam. The H$^{13}$CO$^{+}$ data however trace a peak (L14) which is not prominent in the continuum image at 865 $\mu$m.
It is possible that the W42-MME (L17) compete for material with the nearby sources L13 and L14, which is supported by the potency of MSF in leaf L17 from $M-R$ plot (Figure~\ref{fig5x_ms}) and infall signature in the H$^{12}$CO$^{+}$ spectral profile.
However, the further high-resolution line data with the dense gas modeling can infer the onset of the competitive accretion process \citep{Bonnell2001,Bonnell2006}.
The remaining leaves are observed in the feedback-affected zone, which can be seen in the redshifted velocity range of 66--72 km s$^{-1}$ (Figure~\ref{fig3x}b). Leaves L5 and L2 appear to have potentially accumulated by the outflow emanating from W42-MME (L17).

%

\section{Summary and Conclusion}
\label{sec:conc}

We carried out an analysis of the dense gas structures in the immediate surroundings of a young massive protostar W42-MME using high-resolution (0$''$.31$\times$ 0$''$.25) ALMA dust continuum and line data. A dendrogram analysis is performed in the {\it p-p-v} space, allowing us to trace multi-scale structures and their spatio-kinematic properties. We analyzed the fragmentation and dynamic states of dense structures, considering scales as small as $\sim$2000 AU. The major results of our study are as follows:



\begin{itemize}
 \item The dendrogram analysis of the ALMA H$^{13}$CO$^{+}$ (4--3) data resulted in the identification of 19 dense gas structures, out of which 12 are leaves (mean size $\sim$3780 AU) and 7 are branches (mean size $\sim$11780 AU). 
The gas motions in these structures are transonic to supersonic (1$<\mathcal{M}<5$) and they show overvirial ($\alpha_{\rm vir}\geq2$) states. The infall signature is observed in the H$^{12}$CO$^{+}$ (4--3) line profiles of most of the leaves.

  
\item The non-thermal line-width size relation ($\sigma_{\rm nt}-L$) of dendrogram structures overall shows a weak negative correlation. However, velocity-variation ($\delta\mathit{V_{\rm lsr}}$) displays strong positive correlation with the structure size and the relation follows the generalized Larson's law with steeper exponent of 0.46$\pm$0.14 at a constant column density of $\sim$10$^{23}$ cm$^{-2}$. These results agree with the study of \citet{Ballesteros-Paredes2018} and support the hierarchical and chaotic collapse scenario.


 \item Velocity structure function analysis of H$^{13}$CO$^{+}$ data reveals the strong power-law dependencies ($\propto~L^{0.73\pm0.01}$) with $L \lesssim$6000 AU. 
The overall mean scaling exponents of structure funtion for branch structures is found to be relatively larger compared to that for leaves.

\item The mass-size relation of dendrogram structures shows positive trend with a power-law exponent of 1.73$\pm$0.23. The density, however, displays a marginally statistically significant anti-correlation with size. The leaf L17 that hosts W42-MME, meets the mass-size conditions for MSF as discussed by \cite{Kauffmann2010,Urquhart2014}. 

 

\end{itemize}

Overall, the star formation history in the vicinity of W42-MME appears to have been influenced by the effect of scale dependent physical processes, that include the combined role of turbulence and gravity.


\section[]{Acknowledgments}
We thank the anonymous referee for providing the valuable comments that improved the quality of the paper.
The research work at Physical Research Laboratory is funded by the Department of Space, Government of India. 
L.E.P., A.G.P. and I.I.Z. acknowledge the support of the IAP State Program FFUF-2021-0005.
This research made use of astrodendro, a Python package to compute dendrograms of Astronomical data (http://www.dendrograms.org/).
This paper makes use of the
following ALMA data: ADS/JAO.ALMA\#2018.1.01318.S
and ALMA archive data: ADS/JAO.ALMA\#2019.1.00195.L.
ALMA is a partnership of ESO (representing its member
states), NSF (USA) and NINS (Japan), together with NRC
(Canada), MOST and ASIAA (Taiwan), and KASI (Republic
of Korea), in cooperation with the Republic of Chile. The Joint
ALMA Observatory is operated by ESO, AUI/NRAO and
NAOJ.
This work made use of Astropy:\footnote{http://www.astropy.org} a community-developed core Python package and an ecosystem of tools and resources for astronomy \citep{astropy:2013, astropy:2018, astropy:2022}.

\section{DATA AVAILABILITY}

The ALMA data (IDs: \#2018.1.01318.S and \#2019.1.00195.L) underlying this article are available from the publicly accessible JVO ALMA FITS archive\footnote[1]{http://jvo.nao.ac.jp/portal/alma/archive.do/}.
The ESO VLT/NACO data underlying this article are available from the publicly accessible ESO website\footnote[2]{https://archive.eso.org/eso/eso\_archive\_main.html}.

\bibliography{reference}{}
\bibliographystyle{mnras}


\end{document}